\documentclass[twocolumn,english,aps,pra,10pt,superscriptaddress,floatfix]{revtex4-2}

\usepackage{times}
\usepackage{bm}
\usepackage{graphicx}
\usepackage{amssymb,amsfonts,amsmath,amsbsy,bm,t1enc,latexsym}
\usepackage{float}
\usepackage{siunitx}
\DeclareSIUnit\dBm{dBm}
\DeclareSIUnit\giga{G}
\DeclareSIUnit\baud{Baud}
\usepackage{xcolor}
\definecolor{fancyblue}{RGB}{70, 120, 200}     
\definecolor{fancymagenta}{RGB}{200, 50, 130}  
\usepackage[colorlinks=true,
citecolor=fancyblue,
linkcolor=fancymagenta]{hyperref}

\usepackage[english]{babel}
\usepackage{url}%

\usepackage{color,soul}
\usepackage{textcomp}
\usepackage[]{lipsum}

\usepackage[markup=underlined, defaultcolor=blue]{changes} 
\usepackage{todonotes}

\usepackage{titlesec}
\titlespacing*{\subsection}
{0pt}{0.8em}{0.3em} 

\begin{document}

\title{All-band photonic integrated optical parametric amplification}

\author{Nikolai Kuznetsov}
\thanks{These authors contributed equally to this work.}
\affiliation{Institute of Physics, Swiss Federal Institute of Technology Lausanne (EPFL), CH-1015 Lausanne, Switzerland}

\author{Zihan Li}
\thanks{These authors contributed equally to this work.}
\affiliation{Institute of Physics, Swiss Federal Institute of Technology Lausanne (EPFL), CH-1015 Lausanne, Switzerland}

\author{Tobias J. Kippenberg}
\email[]{tobias.kippenberg@epfl.ch}
\affiliation{Institute of Physics, Swiss Federal Institute of Technology Lausanne (EPFL), CH-1015 Lausanne, Switzerland}
\affiliation{Institute of Electrical and Micro Engineering (IEM), Swiss Federal Institute of Technology Lausanne (EPFL), CH-1015 Lausanne, Switzerland}

\maketitle

\textbf{
Optical amplifiers are ubiquitous in science and technology alike and are the workhorse of modern long-range communications.
Currently, virtually all amplifiers rely on atomic resonances, such as rare-earth-doped fibers, or are based on III-V semiconductors.
Fueled by emerging applications, including co-packaged optics in data centers, neutral-atom-based quantum computing, and ultra-low-loss anti-resonant fibers, there is increased interest in gain beyond traditional wavelength ranges, as well as demand for amplifiers that are high-gain, broadband, low-noise, and deliver high output power.
Over the past few decades, it has been shown that optical parametric amplifiers (OPAs), relying on intrinsic nonlinear response of a material~\cite{hansryd_BroadbandContinuouswavepumpedFiber_2001, hansryd_FiberbasedOpticalParametric_2002, marhic+_FiberOpticalParametric_2015, torounidis_FiberopticalParametricAmplifier_2006}, can address this challenge.
Pioneering works on highly nonlinear optical fibers or bulk crystals have demonstrated the potential of this new class of amplifiers, but high pump powers and long fiber length limited their practical use~\cite{hansryd_BroadbandContinuouswavepumpedFiber_2001, hansryd_FiberbasedOpticalParametric_2002, marhic+_FiberOpticalParametric_2015, torounidis_FiberopticalParametricAmplifier_2006, umeki_HighlyEfficientWavelength_2010, kashiwazaki_HighgainOpticalParametric_2021, kobayashi_WideBandInlineAmplifiedWDM_2021, asobe_BroadbandOpticalParametric_2022}.
Recently, a renaissance of OPAs has occurred with the demonstration of silicon nitride photonic integrated circuits, which exhibit higher effective nonlinearity and enable wider bandwidths~\cite{riemensberger_PhotonicIntegratedContinuoustravellingwave_2022, ye_OvercomingQuantumLimit_2021,kuznetsov_UltrabroadbandPhotonicchipbasedParametric_2025, zhao_UltrabroadbandOpticalAmplification_2025}.
Yet they require ultra-low loss, highly precise dispersion engineering, and large chip footprints, limiting OPA performance to date.
Here we overcome these limitations and, using periodically poled thin-film lithium tantalate (PPLT) photonic integrated circuits~\cite{kuznetsov_WattlevelSecondHarmonic_2025, wang_LithiumTantalatePhotonic_2024}, we demonstrate continuous-wave optical parametric gain up to~\SI{23.5}{\dB}, with a flat-top profile covering more than an~\SI{850}{\nano\meter}-wide optical wavelength window, corresponding to~\SI{100}{\tera\hertz} -- i.e., 20 times wider than the erbium gain bandwidth and covering all communication bands.
Moreover, on-chip output signal power as large as~\SI{313}{\milli\watt} in the optical O-band is achieved.
We further realize all-optical inter-band modulation transfer between the C- and O-bands. 
Our approach uses cascaded second-order nonlinear processes~\cite{umeki_HighlyEfficientWavelength_2010} that overcome the trade-off between low-index, low-nonlinearity photonic platforms and high-index materials with strong Kerr nonlinearity but limited transparency window, by providing effective third-order nonlinearities exceeding $ \gamma $ = \SI{100}{\watt^{-1}\meter^{-1}} while preserving the wide bandgap of the material.
These results establish PPLT integrated photonic circuits, a material already commercially used in RF filters, as a scalable platform for broadband optical amplification and frequency conversion across wavelengths where rare-earth doped amplifiers are absent.
}

Optical amplification has become a backbone of modern information technology, with erbium-doped fiber amplifiers (EDFAs)~\cite{desurvire_HighgainErbiumdopedTravelingwave_1987,mears_LownoiseErbiumdopedFibre_1987} chief among its implementations, enabling long-haul optical communication, global fiber-optic networks, and modern applications ranging from satellite communications to data-center interconnects.
Despite its maturity, there is growing demand for amplification beyond the conventional erbium gain window, driven by multiband S-, C-, and L-band transmission for high-capacity telecommunications, O-band data-center interconnects, low-loss long-wave infrared anti-resonant fibers~\cite{petrovich_BroadbandOpticalFibre_2025}, and neutral-atom quantum technologies operating in the near-infrared and visible.
An ideal optical amplifier would be compact and provide high gain, ultra-broadband unidirectional operation, high saturation power, and quantum-limited noise performance.

Traveling-wave optical parametric amplifiers (OPAs), implemented using the Kerr nonlinear susceptibility $\chi^{(3)}$ in optical fibers and studied extensively over past decades, provide a promising route toward such an ideal.
These devices exhibit unidirectional gain, a large dynamic range, and a quantum-limited noise figure of~\SI{3}{\dB}, with bandwidth fundamentally limited only by dispersion.
Their gain is in situ tunable via the pump power, in contrast to EDFAs and III–V semiconductor amplifiers, whose noise performance degrades under partial inversion.
Collectively, these properties make OPAs strong candidates for next-generation optical amplification.
In particular, they offer the potential to exploit the full bandwidth of current and future fiber systems, increasing wavelength-division multiplexing capacity by more than an order of magnitude~\cite{hansryd_BroadbandContinuouswavepumpedFiber_2001, hansryd_FiberbasedOpticalParametric_2002, torounidis_FiberopticalParametricAmplifier_2006, marhic+_FiberOpticalParametric_2015}.
In addition, OPAs enable wavelength conversion through the associated idler field and support phase-sensitive operation, allowing for noiseless amplification~\cite{hansryd_FiberbasedOpticalParametric_2002, andrekson_FiberbasedPhasesensitiveOptical_2020}.

\begin{figure*}[htb!]
	\centering
	\includegraphics[width=1\textwidth]{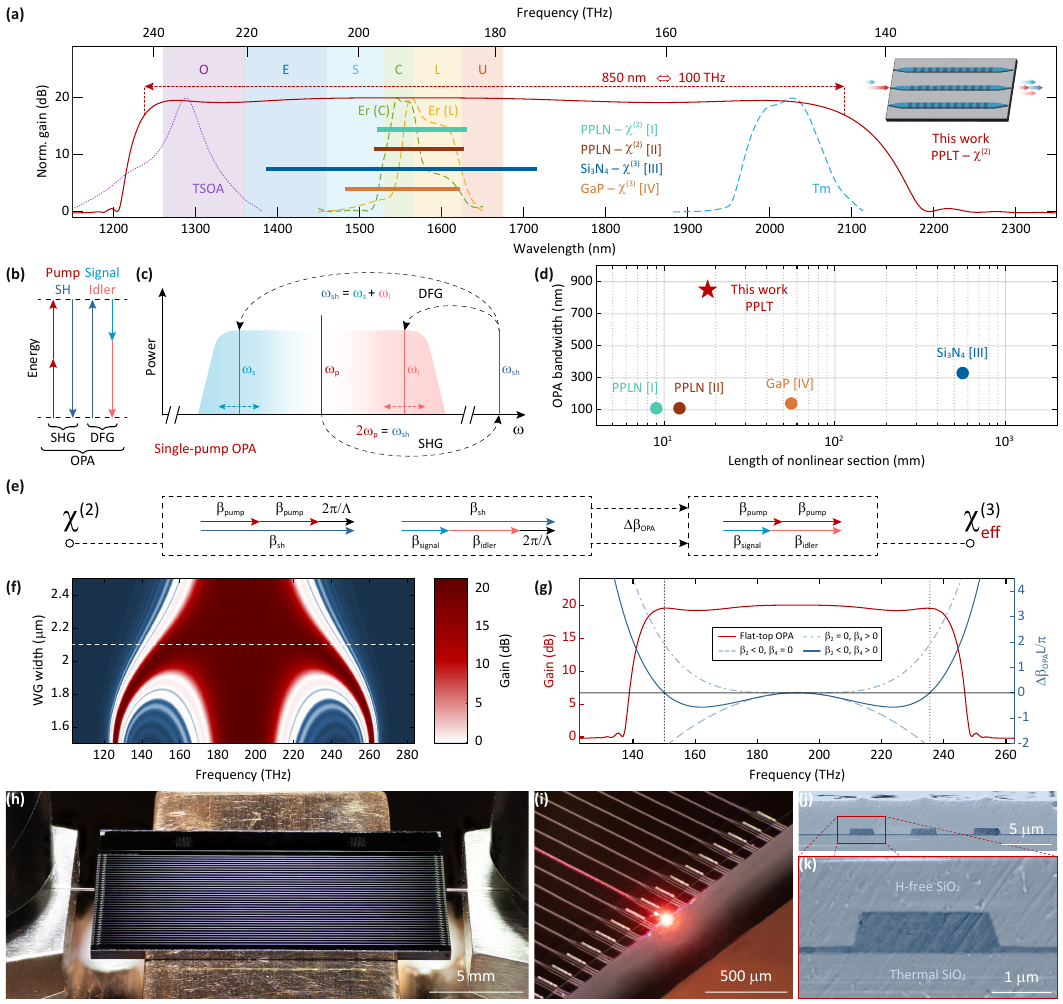}
	\caption{\textbf{Cascaded second-order optical parametric amplification in periodically poled thin-film lithium tantalate (PPLT) waveguides.} 
		\textbf{(a)}~Comparison of the gain bandwidth of the PPLT OPA, demonstrated in this work (solid red line), with other amplifiers.
		A~\SI{3}{\dB}-bandwidth of the PPLT amplifier, reaching~\SI{850}{\nano\meter}, is marked with a red arrow.
        Green and yellow dashed lines -- C-band and L-band erbium-doped fiber amplifiers (EDFAs), respectively.
        Purple dotted line in the O-band -- a tapered semiconductor optical amplifier (TSOA).
        Light-blue dashed line at longer wavelengths -- a thulium-doped fiber amplifier (TDFA).
        All amplification profiles are normalized and meant for bandwidth comparison only; gain values are equalized.
        Colored bars indicate bandwidths of recently demonstrated state-of-the-art photonic integrated OPAs (I -- Ref.~\cite{dean_LowpowerIntegratedOptical_2026}, II -- Ref.~\cite{chen_HighgainOpticalParametric_2025}, III -- Ref.~\cite{zhao_UltrabroadbandOpticalAmplification_2025}, IV -- Ref.~\cite{kuznetsov_UltrabroadbandPhotonicchipbasedParametric_2025}).
        \textbf{(b,~c)}~Energy diagrams and schematic spectra illustrating cascaded parametric amplification in the single-pump~\textbf{(b,~c)} regime.
        In the dual-pump regime, the sum-frequency generation process is used instead of second-harmonic generation.
        \textbf{(e)}~Comparison of bandwidth and nonlinear length of the same photonic integrated OPAs as in~\textbf{(a)} with the PPLT OPA demonstrated in this work.
        \textbf{(e)}~Concept of cascaded second-order parametric amplification, and the quasi-phase-matching (QPM) conditions for second-harmonic generation and difference-frequency generation processes.
        The QPM period is denoted as $\Lambda$.
        \textbf{(f,~g)}~A principle of dispersion engineering, illustrating how, by varying a waveguide width, negative second-order and positive fourth-order dispersion components are combined to yield a broadband flat-top gain profile.
        \textbf{(h)}~A focus-stacked photograph of the photonic chip used in this work.
        Scale bar:~\SI{5}{\milli\meter}.
        \textbf{(i)}~A focus-stacked photograph of the waveguide output with SF emission at~\SI{710}{\nano\meter}, generated by O-band and C-band pumps.
        Scale bar:~\SI{500}{\micro\meter}.
        \textbf{(j,~k)}~False-colored SEM images of cross-sections of three fabricated testing waveguides.
        Scale bars:~\SI{5}{\micro\meter} and~\SI{1}{\micro\meter}, respectively.
	}
	\label{fig:intro}
\end{figure*}

However, fiber-based OPAs remain challenging to deploy in practice, as they require hundreds of meters of highly nonlinear fiber and mitigation of Brillouin scattering due to the high pump powers involved~\cite{torounidis_FiberopticalParametricAmplifier_2006}.
An alternative approach to achieve parametric gain exploits second-order nonlinear processes based on the $\chi^{(2)}$ susceptibility, such as second-harmonic generation (SHG) and difference-frequency generation (DFG) (Fig.~\ref{fig:intro}(b,~c)), and has been demonstrated in periodically poled lithium niobate (LiNbO$_3$, PPLN) ridge waveguides~\cite{umeki_HighlyEfficientWavelength_2010, kashiwazaki_HighgainOpticalParametric_2021, asobe_BroadbandOpticalParametric_2022, kobayashi_WideBandInlineAmplifiedWDM_2021}.
A conventional second-order OPA requires direct pumping at twice the fundamental frequency or a second nonlinear crystal to produce a strong SH wave.

An elegant configuration is to use cascaded $\chi^{(2)}$ nonlinear processes in the same media, that is well-known in bulk crystals~\cite{chou_15mmbandWavelengthConversion_1999,banfi_WavelengthShiftingAmplification_1998,banfi_WavelengthShiftingCascaded_2000,stegeman_2CascadingPhenomena_1996,yablonovitch_AnisotropicInterferenceThreeWave_1972} and was pioneered in ridge waveguides by the NTT laboratories~\cite{umeki_HighlyEfficientWavelength_2010}.
This approach allows to use mature telecom pump sources, including EDFAs, instead of direct pumping by visible lasers.
Quadratic nonlinearity, inherent to non-centrosymmetric crystals, is used in a cascaded process in which a telecom pump photon with the angular frequency $ \omega_{\mathrm{p}} $ first produces near-visible light $  \omega_{\mathrm{sh}} = 2\omega_{\mathrm{p}} $ through the SHG process, which subsequently down-converts through a DFG process into signal and idler photons $  \omega_{\mathrm{sh}} = \omega_{\mathrm{s}} + \omega_{\mathrm{i}}$ (Fig.~\ref{fig:intro}(b,~c)).
This enables $\sim$\SI{e5}{} times higher effective cubic nonlinearity than intrinsic Kerr nonlinearity in glass fibers alone can provide at equivalent pump power and propagation distance.
In ridge waveguides, material dispersion dominates, limiting the bandwidth to approximately~\SI{100}{\nano\meter}~\cite{umeki_HighlyEfficientWavelength_2010, kashiwazaki_HighgainOpticalParametric_2021, kobayashi_WideBandInlineAmplifiedWDM_2021}.

Recently, traveling-wave OPAs have seen a resurgence of interest, with the demonstration of time-continuous net gain in photonic integrated circuits (PICs)~\cite{riemensberger_PhotonicIntegratedContinuoustravellingwave_2022, ye_OvercomingQuantumLimit_2021}.
Beyond significantly higher effective nonlinearity $\gamma$, PICs offer the additional advantage of dispersion engineering through precise control of waveguide geometry.
OPAs based on both Kerr~\cite{riemensberger_PhotonicIntegratedContinuoustravellingwave_2022, ye_OvercomingQuantumLimit_2021, kuznetsov_UltrabroadbandPhotonicchipbasedParametric_2025, zhao_UltrabroadbandOpticalAmplification_2025} and second-order nonlinearities~\cite{chen_HighgainOpticalParametric_2025, dean_LowpowerIntegratedOptical_2026} have recently been shown in various photonic integrated platforms, including silicon nitride and gallium phosphide, showing clear advantages over their fiber-based and ridge-waveguide counterparts~\cite{hansryd_BroadbandContinuouswavepumpedFiber_2001, hansryd_FiberbasedOpticalParametric_2002, marhic+_FiberOpticalParametric_2015, torounidis_FiberopticalParametricAmplifier_2006, umeki_HighlyEfficientWavelength_2010, kashiwazaki_HighgainOpticalParametric_2021, kobayashi_WideBandInlineAmplifiedWDM_2021, asobe_BroadbandOpticalParametric_2022}.
PPLN circuits have been used to extend the bandwidth of frequency conversion in a DFG process~\cite{koyaz_UltrabroadbandTunableDifference_2024, li_IntegratedBroadbandHighefficiency_2025}.
However, fabricating highly nonlinear waveguides or meter-long silicon nitride waveguides is demanding, and lithium niobate has power-handling capability, typically limited to milliwatt range.
Despite significant progress, a short, foundry-compatible PIC-based OPA combining substantially wider bandwidth, high gain, high output power, and quantum-limited noise performance has not yet been demonstrated.

In this work, we address this limitation and demonstrate a photonic integrated OPA providing, on average,~\SI{16}{\dB} of continuous-wave flat-top gain over an unprecedented bandwidth exceeding~\SI{850}{\nano\meter}, which corresponds to~\SI{100}{\tera\hertz} (Fig.~\ref{fig:intro}(a,~d)) -- more than ~\SI{20}{} times the bandwidth of conventional EDFAs.
The maximum gain reaches~\SI{23.5}{\dB}, and we measure~\SI{313}{\milli\watt} of the on-chip output signal power in the O-band.
This is achieved by exploiting the second-order nonlinearity in quasi-phase matched (QPM)~\cite{fejer_QuasiphasematchedSecondHarmonic_1992, wang_UltrahighefficiencyWavelengthConversion_2018} waveguides based on the recently emerged LiTaO$_3$ photonic integrated circuit platform~\cite{wang_LithiumTantalatePhotonic_2024, hulyal_ArrayedWaveguideGratings_2025, wang_UltrabroadbandThinfilmLithium_2024, kuznetsov_WattlevelSecondHarmonic_2025, shelton_RobustPolingFrequency_2025}.
Moreover, we demonstrate all-optical modulation transfer from the optical C-band to the O-band.
These results represent a significant step toward compact, practical, foundry-compatible photonic integrated OPAs, operating in low-noise CW regime, capable of handling high input and output powers, and providing ultra-broadband flat-top gain.
By demonstrating interband modulation transfer, we highlight the potential of PPLT-based OPAs for direct telecom-datacom links, offering a compelling alternative to conventional optical-electrical-optical conversion.

\subsection*{Design and fabrication of periodically poled thin-film lithium tantalate waveguides}

Our approach employs~\SI{18}{\milli\meter}-long periodically poled thin-film lithium tantalate~\cite{kuznetsov_WattlevelSecondHarmonic_2025} (LiTaO$_3$, PPLT) waveguides fabricated on chips (Fig.~\ref{fig:intro}(h-k)) diced from commercially available ion-beam-trimmed wafers (OmedaSemi, and iSABers Group Co., Ltd).
Non-uniformity in periodically poled thin-film ferroelectric waveguides is known to limit nonlinear conversion efficiency~\cite{chen_AdaptedPolingBreak_2024}.
Although this issue can be mitigated using local tuning or adapted poling techniques~\cite{chen_AdaptedPolingBreak_2024}, our approach provides a more robust, scalable, and user-friendly solution; it has already been shown to improve the performance of integrated PPLN OPAs~\cite{chen_HighgainOpticalParametric_2025}.
Periodic poling methods developed in our previous work~\cite{kuznetsov_WattlevelSecondHarmonic_2025} enable uniform poling over extended waveguide lengths.
We use ion-beam-trimmed lithium tantalate wafers with a mean thickness of~\SI{690.25}{\nano\meter} and a variance of~\SI{0.9}{\nano\meter} across the entire 6-inch wafer.
The waveguides are designed based on the results of numerical simulations in the commercially available finite-element-method solver, COMSOL Multiphysics\textsuperscript{\textregistered} (Fig.~\ref{fig:shg}(b--f); see Supplementary Material).
We then fabricate and inspect PPLT waveguides following the methods described in Ref.~\cite{kuznetsov_WattlevelSecondHarmonic_2025}.

To achieve anomalous dispersion, we retain a~\SI{100}{\nano\meter} slab; the designed waveguide width is chosen to be~\SI{1.8}{\micro\meter}.
However, the refractive index data of the bulk material used~\cite{moutzouris_TemperaturedependentVisibleNearinfrared_2011}, together with deviations of the actual waveguide dimensions from the target values, lead to inaccurate dispersion predictions, and we fabricate waveguides with different widths and poling periods on the same dies (Fig.~\ref{fig:shg}(e,~f)).
We find waveguides providing the required QPM and dispersion properties by introducing a~\SI{40}{\nano\meter} offset on the poling period and a~\SI{300}{\nano\meter} offset on the waveguide width compared to the original parameters.
The optimized design has a~\SI{5305}{\nano\meter} poling period and a~\SI{2.1}{\micro\meter} waveguide width (Fig.~\ref{fig:shg}(b)).
We proceed with the fabrication of subsequent samples using these experimentally verified dimensions.
The total length of the waveguides is~\SI{20}{\milli\meter}, and the length of the periodically poled section is~\SI{18}{\milli\meter}.
In principle, our fabrication methods and poling techniques can be extended to longer waveguides in the future to further increase the efficiency and reduce pump power requirements.
\begin{figure*}[htb!]
	\centering
	\includegraphics[width=1\textwidth]{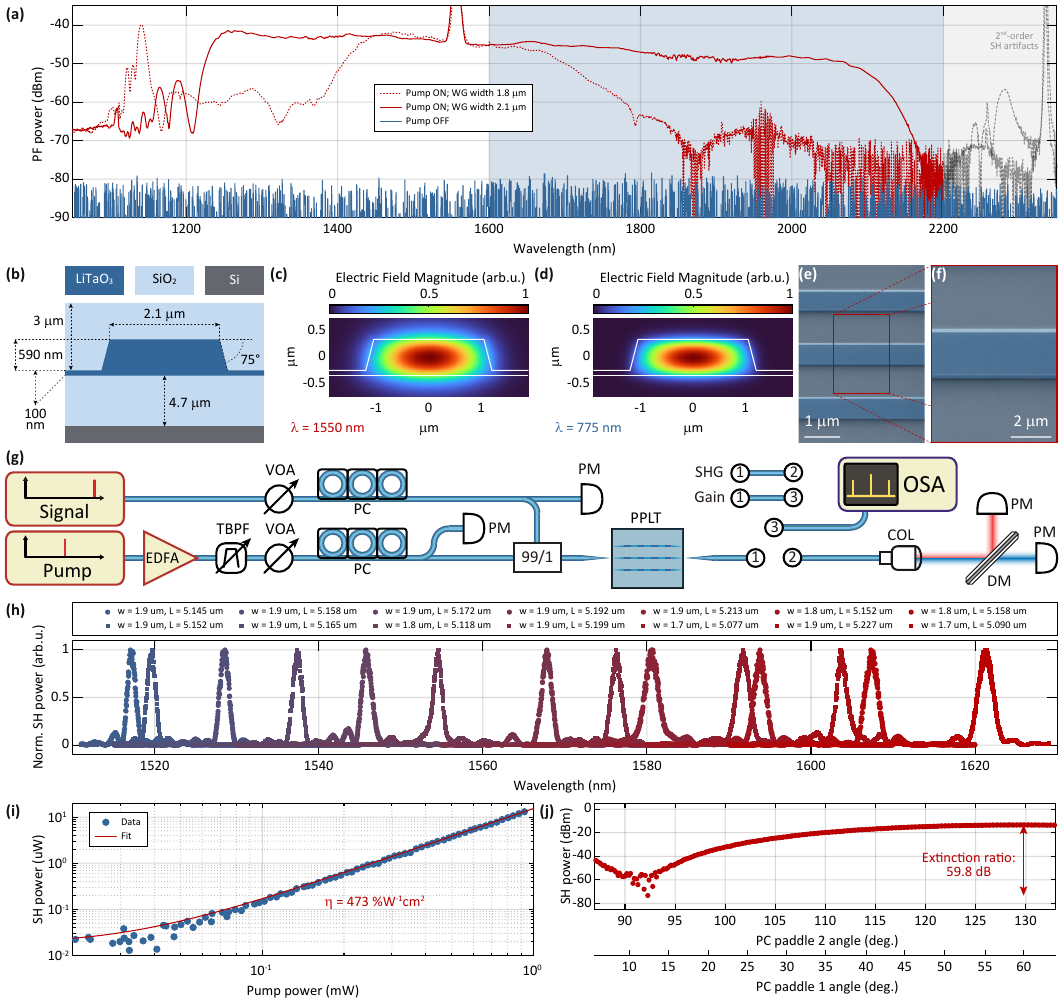}
    \caption{\textbf{Second-harmonic generation in long periodically poled lithium tantalate waveguides.}
        \textbf{(a)}~Optical parametric fluorescence spectra in waveguides with~\SI{1.8}{\micro\meter} (dotted line)~and~\SI{2.1}{\micro\meter}~\textbf{(k)} (solid line)~widths, measured at~\SI{1}{\watt} of pump power.
        The blue-shaded region indicates the data measured with the long-range OSA.
        The gray-shaded region indicates ghost lines generated by higher-order scattering of shorter-wavelength signals in the OSA grating.
        Specifically, the strongest rightmost peak is the third-order scattering of the SH signal.
        \textbf{(b)}~A schematic of the designed waveguide cross-section, showing the material stack and the target dimensions.
        \textbf{(c,~d)}~Numerically calculated fundamental mode profiles of the pump and SH waves, respectively.
        \textbf{(e,~f)}~False-colored top-view SEM images of three fabricated waveguides taken before cladding deposition.
        Scale bars:~\SI{1}{\micro\meter} and~\SI{2}{\micro\meter}, respectively.
        \textbf{(g)}~A schematic of the experimental setup used to characterize the SHG and OPA processes in PPLT waveguides.
        EDFA: erbium-doped fiber amplifier.
        TBPF: tunable bandpass filter.
        VOA: variable optical attenuator.
        PC: polarization controller.
        PM: power meter.
        COL: collimator.
        DM: dichroic mirror.
        OSA: optical spectrum analyzer.
        \textbf{(h)}~Examples of normalized SH spectra measured in waveguides with varying widths and poling periods.
        The spectral profiles show negligible distortion from the canonical shape, indicating high uniformity of the waveguide dimensions.
        \textbf{(i)}~Measured low-power SHG efficiency and high-power SH conversion.
        The efficiency curve fitting suggests an efficiency of~\SI{473}{\percent\watt^{-1}\centi\meter^{-2}}.
        \textbf{(j)}~Polarization dependence of the SHG process, measured by synchronously varying the angles of two paddles of a motorized polarization controller.
        The polarization extinction approaches~\SI{60}{\dB}.        
	}
	\label{fig:shg}
\end{figure*}

We measure the SHG using the setup shown in Fig.~\ref{fig:shg}(g), keeping the pump laser (TOPTICA CTL) power low and undepleted at~\SI{1}{\milli\watt}.
In all experiments in this work, we use lensed fibers for input and output coupling to PPLT waveguides.
The average fiber-to-chip insertion loss is~\SI{2.1}{\dB} in the optical C-band, and~\SI{3.7}{\dB} in the O-band.
Fig.~\ref{fig:shg}(h) shows the normalized SHG spectra measured in different waveguides with varying widths and poling periods.
While minor spectral distortions are present, each spectrum exhibits a single dominant SHG peak, as expected for an ideal uniform waveguide.
The measured normalized SHG efficiency is~\SI{473}{\percent\watt^{-1}\centi\meter^{-2}} (Fig.~\ref{fig:shg}(i)), which is not far from the theoretical value of~\SI{617}{\percent\watt^{-1}\centi\meter^{-2}} based on the material nonlinearity d$_{33}$
reported as~\SI{10.7}{\pico\meter\volt^{-1}} in Ref.~\cite{shoji_AbsoluteScaleSecondorder_1997}.
We note that this value is reported for~\SI{1313}{\nano\meter} wavelength, and the actual value at C-band might be lower.
By measuring the SHG polarization extinction ratio (Fig.~\ref{fig:shg}(j)), we ensure that interacting modes are highly polarized and frequency conversion is not affected by higher-order modes.

We use parametric fluorescence measurements to estimate the amplification and frequency conversion bandwidth and select waveguides for direct gain measurements.
While previous demonstrations of broadband frequency conversion rely on system tunability to access different spectral regions sequentially~\cite{asobe_BroadbandOpticalParametric_2022, koyaz_UltrabroadbandTunableDifference_2024}, our design operates in the optimized dispersion regime.
The contribution of the Kerr nonlinear phase shift is negligible due to the short interaction length.
Here, the combination of negative second-order dispersion $ \beta_{4} $ and positive fourth-order dispersion $ \beta_{4} $ enables flat broadband operation with a fixed pump wavelength
\begin{equation*}
	\Delta \beta_{\mathrm{opa}} \approx \beta_{2}(\omega_{\mathrm{p}} - \omega_{\mathrm{s}}))^{2} + \frac{1}{12}\beta_{4}(\omega_{\mathrm{p}} - \omega_{\mathrm{s}})^{4} + \cdots.
	\label{eq:taylor_pm_main}
\end{equation*}
Here, $ \omega_{\mathrm{p}} $ and $ \omega_{\mathrm{s}} $ are pump and signal angular frequencies, respectively (Fig.~\ref{fig:intro}(b,~c)).
The output of the setup shown in Fig.~\ref{fig:shg}(g) is directed to one of two OSAs (Yokogawa AQ6370D and Yokogawa AQ6375) with~\SI{2}{\nano\meter} resolution bandwidth, depending on the wavelength range of interest.
Fig.~\ref{fig:shg}(a) shows the parametric fluorescence spectra obtained under~\SI{1}{\watt} of pump power in waveguides with~\SI{1.8}{\micro\meter} and~\SI{2.1}{\micro\meter} widths, respectively.
The first spectrum (dotted line) features the central gain band and the short-wavelength gain sideband originating from fourth-order dispersion.
The long-wavelength counterpart lies beyond the measurement range and is expected to be located at approximately~\SI{2500}{\nano\meter}.
The second plot (solid line) represents a flat-top spectrum spanning from nearly ~\SI{1200}{\nano\meter} to almost~\SI{2200}{\nano\meter}.
This bandwidth is slightly wider than predicted by numerical simulations based on the sample thickness and the designed waveguide cross-section, and by fitting the gain profile, we estimate the dispersion coefficients as $ \beta_2 \approx  $~\SI{-5.2}{\femto\second^2\milli\meter^{-1}} and $ \beta_4 \approx  $~\SI{775}{\femto\second^4\milli\meter^{-1}}.

\subsection*{Optical parametric amplification via cascaded second-order nonlinear conversion}

\begin{figure*}[htb!]
	\centering
	\includegraphics[width=1\textwidth]{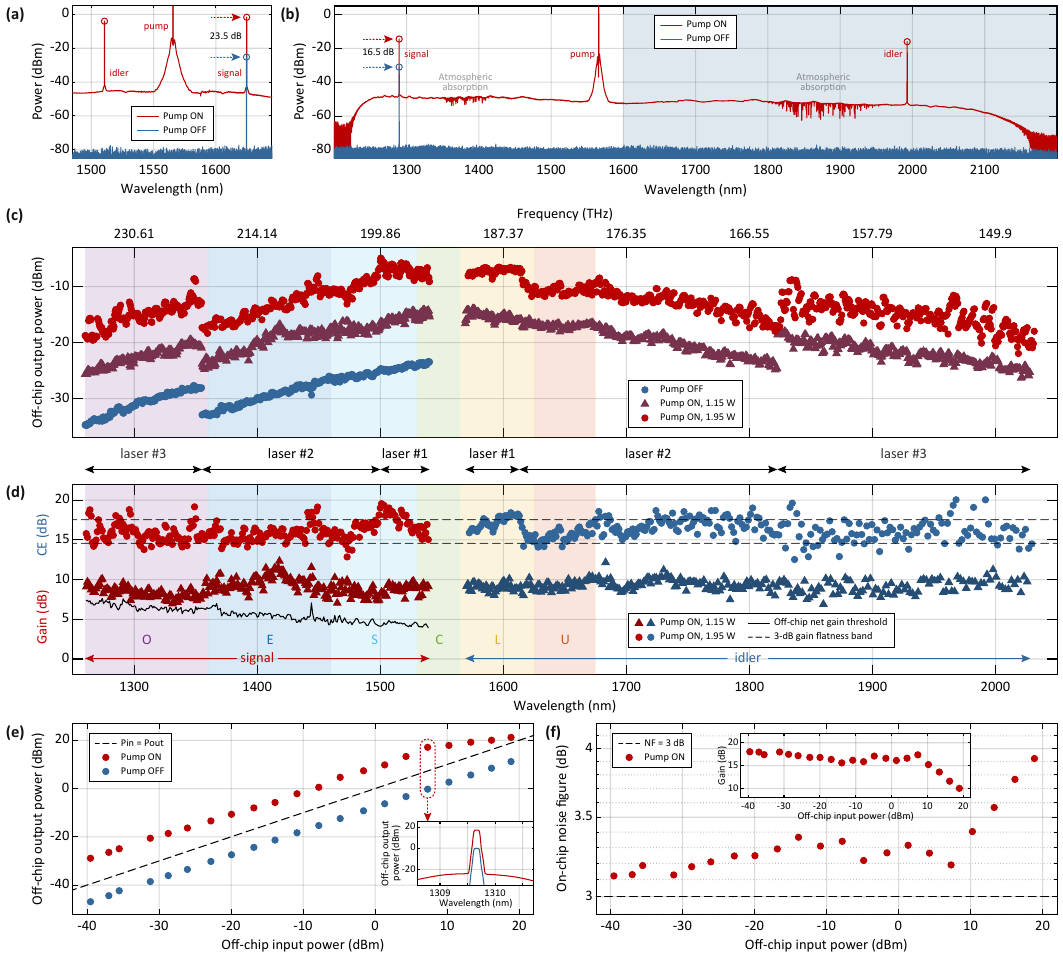}
    \caption{\textbf{Measurements of broadband cascaded amplification in thin-film PPLT waveguides.} 
		\textbf{(a)}~Single-frequency gain measurement.
        The signal wavelength is~\SI{1625}{\nano\meter}.
        The blue line shows the transmission spectrum with the pump switched off, and the red line shows the amplified spectrum.
        The on-off gain is~\SI{23.5}{\dB}, and the pump power is~\SI{2.3}{\watt}.
        \textbf{(b)}~Same as~\textbf{(a)}, but with a signal wavelength of~\SI{1290}{\nano\meter}.
        The on-off gain is~\SI{16.5}{\dB}, and the pump power is~\SI{1.95}{\watt}.
        The blue-shaded region is measured with the long-range OSA.
        An idler wave is generated in a remote band at~\SI{1992}{\nano\meter}.
        \textbf{(c)}~Hyper-wideband gain measured with an OSA in the "max-hold" mode using three frequency-swept lasers over their entire available frequency ranges, marked with the arrows.
        The idler data, consisting of all data points at wavelengths above~\SI{1550}{\nano\meter}, are measured with the long-range OSA.
        \textbf{(d)}~Gain and conversion efficiency calculated from the data shown in~\textbf{(c)}.
        \textbf{(e,~f)}~Output signal power and noise figure, respectively, measured for different input signal power levels.
        The maximum off-chip output signal power is~\SI{132}{\milli\watt}, corresponding to~\SI{313}{\milli\watt} of on-chip signal power.
        The inset in~\textbf{(e)} shows optical spectra for the marked data points.
        The inset in~\textbf{(f)} shows the calculated gain versus input signal power.
	}
	\label{fig:opa}
\end{figure*}

We measure the single-frequency gain and conversion efficiency for a signal at \SI{1630}{\nano\meter} in the undepleted regime (Fig.~\ref{fig:opa}(a)) using the same methods as described in our previous works~\cite{riemensberger_PhotonicIntegratedContinuoustravellingwave_2022, kuznetsov_UltrabroadbandPhotonicchipbasedParametric_2025}.
We combine the pump amplified by an EDFA (Keopsys~CEFA-C) and signal waves using a \SI{99}{}/\SI{1}{} fiber coupler.
After careful coupling optimization, we measure~\SI{23.5}{\dB} of on-off gain at \SI{2.3}{\watt} of off-chip pump power, corresponding to \SI{18}{\dB} of fiber-to-fiber gain (Fig.~\ref{fig:opa}(a), \SI{0.1}{\nano\meter} resolution bandwidth).
For the waveguide with the length of the nonlinear section equal to $ \mathrm{L} $, gain $ G_{\textrm{s}} $ and conversion efficiency $ G_{\textrm{i}} $ are defined, respectively, as
\begin{equation*}
	G_{\textrm{s}} = \dfrac{P_{\textrm{s}}(\mathrm{L})}{P_{\textrm{s}}(0)}, \quad
	G_{\textrm{i}} = \dfrac{P_{\textrm{i}}(\mathrm{L})}{P_{\textrm{s}}(0)},
\end{equation*}
where $ P_{\textrm{s}} $ and $ P_{\textrm{i}} $ are signal and idler powers, respectively.
Considering degenerate four-wave mixing, the equivalent Kerr nonlinearity for a waveguide with the same experimental conditions ($ \mathrm{L} $ = \SI{18}{\milli\meter}, $ A_{\mathrm{eff}} $ = \SI{1.37}{\micro\meter^2}, $ P_{\textrm{p}} $  = \SI{1.42}{\watt} on-chip) would be $ \gamma $ = \SI{133}{\watt^{-1}\meter^{-1}}, as estimated by $ 	G_{\textrm{s}} = 1 +   \sinh(\gamma P_{\textrm{p}} \mathrm{L})^2  $~\cite{hansryd_FiberbasedOpticalParametric_2002}, and nonlinear refractive index $  n_2 $ = \SI{4.5e-17}{\meter^2\watt^{-1}}, which exceeds silicon nitride by two orders of magnitude~\cite{riemensberger_PhotonicIntegratedContinuoustravellingwave_2022} and is almost on par with gallium phosphide~\cite{kuznetsov_UltrabroadbandPhotonicchipbasedParametric_2025}.

Due to the photorefractive effect and reflections from the chip facets, SH generation becomes unstable at this power level, and we observe gain fluctuations and unstable spontaneous generation of OPO sidebands.
Although it is possible to reduce photorefractive fluctuations by heating the sample above~\SI{100}{\celsius}, reflections from the chip facets currently prevent us from using higher pump powers.

Next, similarly, we measure the single-frequency gain for a signal at~\SI{1290}{\nano\meter} (Fig.~\ref{fig:opa}(b),~\SI{0.1}{\nano\meter} resolution bandwidth).
For this measurement, we replace the fiber coupler with a C-band/O-band wavelength-division multiplexer (WDM).
The pump power is set to~\SI{1.95}{\watt}, and the measured on-off gain is~\SI{16.5}{\dB}.
As expected, the idler wave is generated according to energy conservation at a remote wavelength of~\SI{1992}{\nano\meter}.
To experimentally cover the widest possible bandwidth, we install a \SI{90}{}/\SI{10}{} fiber coupler and connect widely tunable lasers covering the wavelength range from~\SI{1260}{\nano\meter} to~\SI{1540}{\nano\meter}, while setting the OSAs to the "max-hold" mode with~\SI{2}{\nano\meter} resolution bandwidth.
We wait until each laser completes several full-range scans to cover all available wavelengths without gaps, and then repeat the procedure for the idler measurements.
The raw measured data are shown in Fig.~\ref{fig:opa}(c), and Fig.~\ref{fig:opa}(d) shows the calculated gain and conversion efficiency.
Although we use lower pump power for stability, signal coupling and polarization inevitably fluctuate over the wide spectral range, resulting in some gain variations.
Despite these fluctuations, the gain profile remains relatively flat -- a notable advantage over Kerr-based OPAs~\cite{hansryd_FiberbasedOpticalParametric_2002, kuznetsov_UltrabroadbandPhotonicchipbasedParametric_2025, zhao_UltrabroadbandOpticalAmplification_2025}.

To verify the performance of the amplifier at high input signal powers, we again replace the fiber coupler with a WDM and connect the O-band signal laser to a TSOA (TOPTICA BoosTA pro) and a DVOA (OZ OPTICS DA-100).
We set the pump power to~\SI{2}{\watt} and measure single-frequency gain spectra over a~\SI{5}{\nano\meter} span around the signal, with a~\SI{0.1}{\nano\meter} resolution bandwidth.
Fig.~\ref{fig:opa}(e) shows the results of these measurements.
Using bandwidth-corrected measurements of the parametric fluorescence power density, we calculate the noise figure for the same data points (Fig.~\ref{fig:opa}(f)).
Saturation begins at approximately~\SI{10}{\milli\watt} of off-chip input signal power.
The maximum output power reaches~\SI{132}{\milli\watt}, corresponding to~\SI{313}{\milli\watt} on-chip after accounting for~\SI{3.7}{\dB} of coupling loss.
The on-chip noise figure is close to the~\SI{3}{\dB} limit of phase-insensitive amplification, as expected, and increases after saturation.

\subsection*{Dual-pump optical parametric amplification}

\begin{figure*}[htb!]
	\centering
	\includegraphics[width=1\textwidth]{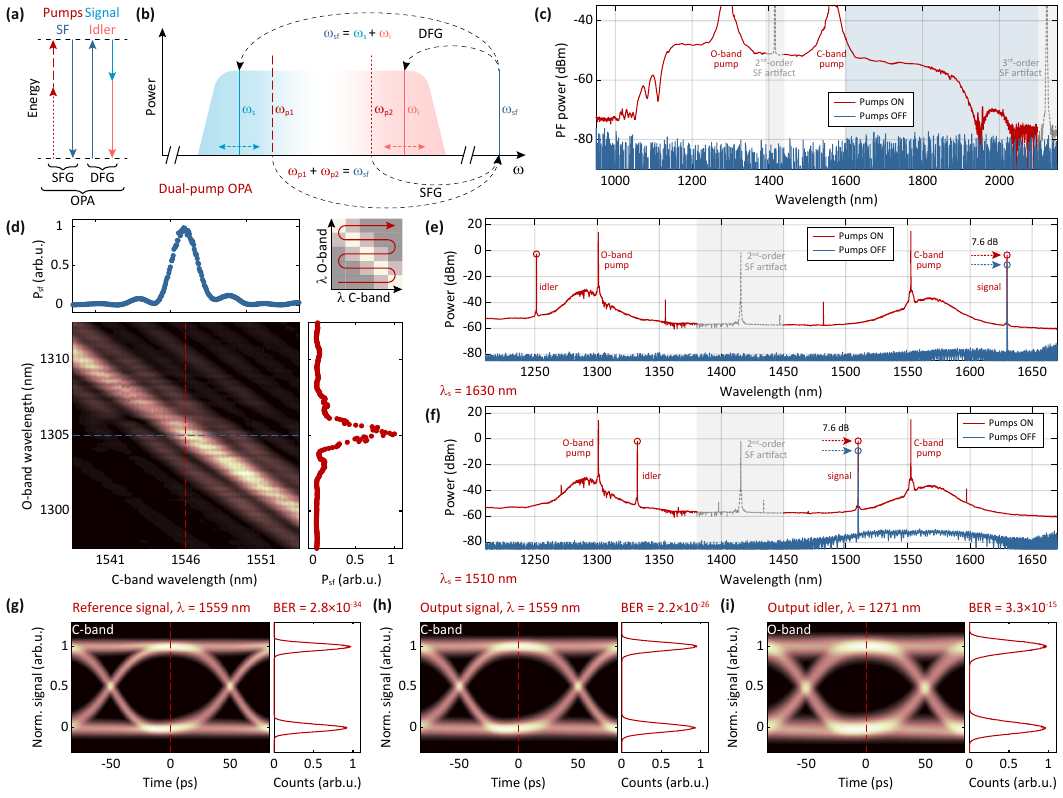}
	\caption{\textbf{Sum-frequency generation, dual-pump optical parametric amplification, and inter-band all-optical modulation transfer.} 
		\textbf{(a,~b)}	Energy diagrams and schematic spectra illustrating cascaded parametric amplification in the dual-pump~\textbf{(b,~c)} regime.
		The sum-frequency generation process is used instead of second-harmonic generation.
      	\textbf{(c)}~Optical parametric fluorescence spectra in a waveguide with a width of~\SI{1.8}{\micro\meter}.
        \textbf{(d)}~An SFG map measured by sweeping two lasers located in the C-band and the O-band.
        The top and side panels show the SFG spectrum of one swept laser at a fixed wavelength of the other.
        A schematic in the top-right corner shows the algorithm of the dual-laser raster scan used for the SFG measurements.
        \textbf{(e,~f)}~Dual-pump amplification spectra for single-frequency signals set to~\SI{1630}{\nano\meter} and~\SI{1510}{\nano\meter}, respectively.
        \textbf{(g--i)}~Eye diagrams obtained for the reference signal before the chip, the output amplified signal, and the output idler, respectively.
        The idler is generated by placing the signal at~\SI{1596}{\nano\meter}.
	}
	\label{fig:dualpump}
\end{figure*}

As mentioned earlier, it is possible to use SFG instead of SHG for cascaded second-order amplification (Fig.~\ref{fig:dualpump}(a,~b).
We use two pump lasers located in the O-band and the C-band to measure the two-dimensional SFG map, evaluate the optical gain, and demonstrate interband all-optical modulation transfer, similar to the demonstrations in Refs.~\cite{kuznetsov_UltrabroadbandPhotonicchipbasedParametric_2025, zhao_UltrabroadbandOpticalAmplification_2025, ye_OvercomingQuantumLimit_2021, kong_SuperbroadbandOnchipContinuous_2022, li_IntegratedBroadbandHighefficiency_2025}.
In this cascaded amplification scheme, the SFG process effectively plays the role of SHG, with a virtual pump located at twice the SFG wavelength, and the principles of the subsequent DFG process remain the same as in the single-pump case.
However, in the specific case of O-band and C-band pumping, the effective single-pump wavelength is around~\SI{1420}{\nano\meter}, in the optical E-band; this is a relatively uncommon wavelength range, and no well-established high-power lasers or amplifiers are available to serve as a single pump source.
For all dual-pump measurements, we use a waveguide with a width of~\SI{1.8}{\micro\meter}, which exhibits excessive anomalous dispersion in the C-band, as shown in Fig.~\ref{fig:shg}(a), but provides suitable dispersion for the virtual single pump at approximately~\SI{1415}{\nano\meter}.
The O-band pump laser is amplified using a TSOA, while the C-band pump laser is amplified using a conventional EDFA; these amplifiers are not used in the low-power SFG map measurement shown in Fig.~\ref{fig:dualpump}(d) (see Supplementary Material).
The parametric fluorescence measurement is shown in Fig.~\ref{fig:dualpump}(c), highlighting a flat-top bandwidth exceeding~\SI{800}{\nano\meter}.

To measure the SFG map, we program the pump lasers to perform a stepwise raster scan, changing the wavelength in~\SI{0.1}{\nano\meter} steps over a~\SI{12}{\nano\meter} range.
At each step, the power meter records the SFG power.
O-band and C-band SFG slices at fixed wavelengths of the companion laser are shown in the top and right panels of Fig.~\ref{fig:dualpump}(d), respectively.
We set the pump wavelengths to~\SI{1299}{\nano\meter} and~\SI{1554}{\nano\meter} for high-power measurements; both pump lasers are then fine-tuned to maximize the SFG power.
The pump waves are combined using a WDM, and the signal is subsequently combined with both pumps using a \SI{90}{}/\SI{10}{} fiber coupler.
The maximum available pump power in the O-band is~\SI{240}{\milli\watt}, limiting the on-off optical gain to~\SI{7.6}{\dB}, as shown in Fig.~\ref{fig:dualpump}(e,~f).
Accordingly, the C-band pump power is set to~\SI{500}{\milli\watt} -- higher than the O-band pump power to ensure the gain is limited by the O-band pump.
Further increasing the C-band pump power does not increase gain, as the SFG process requires one O-band photon for each C-band photon.
Contrary to dual-pump Kerr amplification, where additional idler waves are created due to degenerate FWM processes involving individual pumps, in the cascaded second-order amplification periodic poling is designed only for the SFG process, and individual pumps are poorly phase-matched with their respective SH waves~\cite{hefti_HighpurityFrequencydegeneratePhoton_2025}, and individual DFG processes are suppressed.
Small additional idler lines arising due to FWM processes weak and second-order DFG processes through weak individual SH waves can still be observed in Fig.~\ref{fig:dualpump}(e,~f).
However, these lines are suppressed by nearly~\SI{40}{\dB} compared to the main signal and idler, and their contribution to the signal distortion is below the pump ASE noise leakage.

\subsection*{Telecom-to-datacom all-optical modulation transfer}

A distinctive feature of OPAs is the generation of an idler wave, arising from energy and momentum conservation in the three- and four-wave mixing processes, enabling coherent modulation transfer across different optical bands.
A key feature of dual-pump parametric amplification is that both the signal and the idler can be located in practically relevant, well-established optical bands, together with two strong pumps.
In our experiment, we modulate a signal at~\SI{1596}{\nano\meter}.
This wavelength, formally located in the L-band, is chosen to ensure that the idler can be filtered using the available FBG, which has a limited tunability range of~\SI{6}{\nano\meter} centered at~\SI{1270}{\nano\meter}.
For the reference and amplified signal measurements, we set the wavelength to~\SI{1559}{\nano\meter} to match the available FBG range.
We modulate the signal, preamplified with an L-band EDFA (Keopsys CEFA-L), using a bulk intensity modulator (Optilab IML-1550-40-PM-V-HER) driven by an AWG (Keysight M8195A), generating a~\SI{10}{\giga\baud} NRZ PRBS15 sequence, and separately measure the modulated signal transmitted through the PPLT amplifier and the generated idler using a fast photodiode (Coherent XPDV2320R-VM-FA) and a high-speed oscilloscope (Teledyne LeCroy SDA 8330HD).
Using FBGs and bandpass filters composed of an additional FBG and a WDM, we ensure that all spectral components except the one being measured, including residual pumps, are suppressed by at least~\SI{40}{\dB} (see Supplementary Material).
Fig.~\ref{fig:dualpump}(g--i) shows the acquired eye diagrams for both measurements, along with the reference measurement of the modulated signal before coupling into the PPLT OPA.
Given the nearly instantaneous nature of parametric processes, there are virtually no bandwidth limitations on the transferred modulation, and this demonstration paves the way for direct all-optical telecom-datacom communication links.

\subsection*{Discussion}

By using thin-film PPLT waveguides, we demonstrate cascaded second-order CW optical parametric amplification with a massive flat-top bandwidth reaching~\SI{850}{\nano\meter}.
In addition to the general advantages of parametric amplifiers -- low-noise operation, nearly instantaneous unidirectional gain, and phase-coherent idler generation -- our system leverages several unique aspects of its design and fabrication.
First, ion-beam-trimmed wafers simplify scalability and eliminate the need for complex adapted~\cite{chen_AdaptedPolingBreak_2024} or tunable schemes~\cite{li_AdvancingLargescaleThinfilm_2024} to compensate for film thickness non-uniformity.
Our waveguides are only~\SI{2}{\centi\meter} long and are designed to operate in the anomalous dispersion regime, enabling a continuously broad bandwidth without the need to tune phase matching by adjusting temperature or pump wavelength.
The low material birefringence facilitates anomalous dispersion in silicon dioxide-cladded waveguides and prevents mode crossings.
Moreover, the dispersion design is tolerant to variations in the waveguide cross-section, relaxing fabrication precision requirements.
In contrast to narrowband OPAs~\cite{hansryd_FiberbasedOpticalParametric_2002, kashiwazaki_HighgainOpticalParametric_2021} in optical fibers and bulk crystals, which in practice cannot be arbitrarily positioned across the spectrum and must remain near bands with available high-power pump sources, our PPLT OPA covers all standard communication bands and extends far beyond them, eliminating the need for high-power sources outside mature wavelength regions.
While PPLN PICs have higher nonlinearity~\cite{chen_HighgainOpticalParametric_2025, dean_LowpowerIntegratedOptical_2026}, low birefringence and higher damage threshold~\cite{yan_HighOpticalDamage_2020, suntsov_OpticalDamageResistant_2024, kuznetsov_WattlevelSecondHarmonic_2025} of PPLT PICs allow broadband operation~\cite{zhang_UltrabroadbandIntegratedElectrooptic_2025} with higher output power.
Finally, PPLT OPA enables direct interband modulation transfer, opening a pathway towards all-optical telecom-datacom links.

\footnotesize
\noindent \textbf{Author contributions:}
N.K. developed the concept, performed numerical simulations, developed the periodic poling routines, and carried out SHG and optical gain measurements.
Z.L. developed and optimized fabrication routines for the poling electrodes and the optical layer, and fabricated the samples.
N.K. prepared the manuscript with contributions from all authors. 
T.J.K supervised the work.

\noindent \textbf{Funding information:}
This work has received funding from the European Research Council (ERC) under the Horizon Europe research and innovation programme, grant agreement No.~101167540 (ATHENS).
This work is supported by the EU Horizon Europe EIC programme under grant agreement No.~101187515 (ELLIPTIC ), and by the Swiss State Secretariat for Education, Research and Innovation (SERI).

\noindent \textbf{Acknowledgments:}
We thank Johann Riemensberger for early suggestions for this project.
The samples were fabricated in the EPFL Center of MicroNanoTechnology (CMi) and the Institute of Physics (IPHYS) cleanroom.
Two-photon microscopy imaging was performed in the UNIL Cellular Imaging Facility (CIF).

\noindent \textbf{Disclosures:}
All authors declare no competing interests.

\noindent \textbf{Data availability:}
All experimental datasets and scripts used to produce the plots in this work will be uploaded to the Zenodo repository upon publication of this preprint.

\bibliography{bibliography}

\end{document}


\title{Supplementary Material to: All-band photonic integrated optical parametric amplification}

\author{Nikolai Kuznetsov}
\thanks{These authors contributed equally to this work.}
\affiliation{Institute of Physics, Swiss Federal Institute of Technology Lausanne (EPFL), CH-1015 Lausanne, Switzerland}

\author{Zihan Li}
\thanks{These authors contributed equally to this work.}
\affiliation{Institute of Physics, Swiss Federal Institute of Technology Lausanne (EPFL), CH-1015 Lausanne, Switzerland}

\author{Tobias J. Kippenberg}
\email[]{tobias.kippenberg@epfl.ch}
\affiliation{Institute of Physics, Swiss Federal Institute of Technology Lausanne (EPFL), CH-1015 Lausanne, Switzerland}
\affiliation{Institute of Electrical and Micro Engineering (IEM), Swiss Federal Institute of Technology Lausanne (EPFL), CH-1015 Lausanne, Switzerland}

\maketitle

{\hypersetup{linkcolor=black}\tableofcontents}

\newpage

\renewcommand{\thefigure}{S\arabic{figure}}
\renewcommand{\theequation}{S\arabic{equation}}

\section{Highly uniform ion-beam-trimmed thin-film lithium tantalate wafers}

It has been previously shown that the efficiency of second-harmonic generation and other quasi-phase-matched nonlinear processes in long ferroelectric waveguides is strongly affected by the thickness non-uniformity of thin-film substrates~\cite{chen_AdaptedPolingBreak_2024}.
Thickness variations lead to spatial fluctuations of the effective refractive index and dispersion, resulting in phase-mismatch accumulation along the propagation direction and a corresponding reduction in optical frequency conversion efficiency.

Several approaches have been proposed to mitigate this limitation, including adapted periodic poling~\cite{chen_AdaptedPolingBreak_2024} and local phase-matching tuning using integrated heater arrays~\cite{li_AdvancingLargescaleThinfilm_2024}.
While effective, these techniques introduce additional fabrication complexity, increase process time, and may reduce scalability.
They also require high-resolution thickness metrology and accurate numerical models to predict local phase-matching conditions across the wafer.

In this work, we instead address the problem at the substrate level by using commercially available ion-beam-trimmed thin-film lithium tantalate wafers (OmedaSemi, and iSABers Group Co., Ltd).
The ion-beam trimming process enables post-bonding planarization of the thin film, resulting in significantly improved thickness uniformity compared to standard thin-film lithium tantalate substrates.

\begin{figure*}[htb]
	\centering
	\includegraphics[width=1\textwidth]{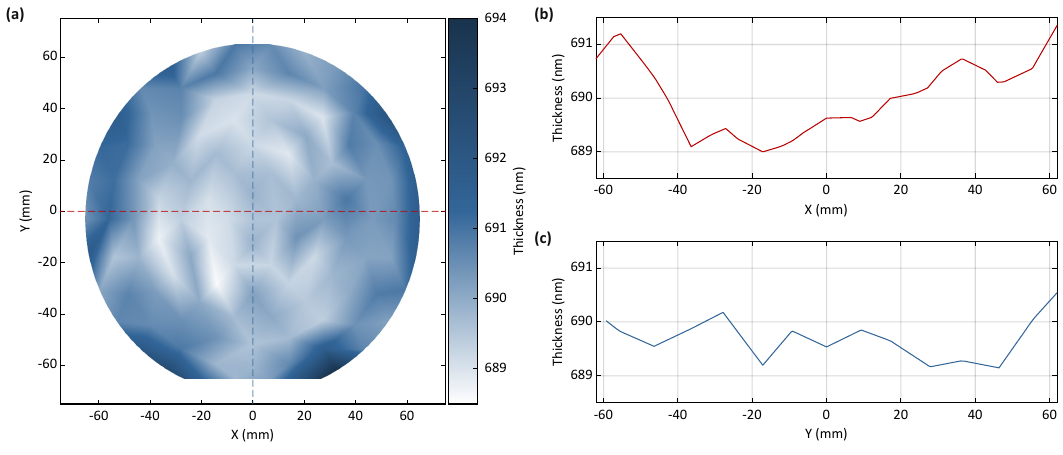}
	\caption{\textbf{Thickness variations of the ion-beam-trimmed lithium tantalate wafer used in this work.}
		\textbf{(a)} Thickness map of the wafer.
        \textbf{(b-c)} Linearly interpolated thickness profiles of the wafer measured at zero coordinates along X and Y axes, respectively.
		}
	\label{fig:thickness_map}
\end{figure*}

The thickness uniformity of the wafer is characterized using optical interferometer (FilMetrics F20-UV) prior to fabrication of photonic circuits.
The resulting thickness map and thickness profiles along zero coordinates are shown in Fig.~\ref{fig:thickness_map}.

The average thin-film thickness is measured to be~\SI{690.25}{\nano\meter}, which allows for achieving the anomalous dispersion required for wide OPA bandwidth.
The peak-to-peak thickness variation across the wafer is~\SI{0.90}{\nano\meter}, representing an order-of-magnitude improvement compared to typical thin-film lithium tantalate substrates.

This level of thickness uniformity ensures that the phase-matching condition remains nearly constant over centimeter-scale propagation lengths, eliminating the need for post-fabrication phase-matching compensation and enabling reproducible fabrication of long, high-efficiency nonlinear waveguides.

We dice the trimmed wafers into smaller dies prior to electron-beam lithography and subsequent processing.
Chip-scale fabrication is chosen to allow tighter process control, faster iteration cycles, and reduced material consumption during development.
Our fabrication flow is, however, fully compatible with wafer-scale processing.

To suppress leakage current, a~\SI{100}{\nano\meter} silicon dioxide layer is first deposited by plasma-enhanced chemical vapor deposition (PlasmaLab 100).
Poling electrodes are then defined using \SI{100}{\kilo\volt} electron-beam lithography (EBL, Raith EBPG5000) with a bilayer resist (PMMA/MMA).
Subsequently, a \SI{100}{\nano\meter} aluminum layer is deposited by electron-beam evaporation (Alliance-Concept EVA 451), followed by lift-off in acetone to remove excess metal.
Prior to high-voltage pulse poling, a~\SI{2}{\micro\meter} photoresist layer (ECI 3027) is spin-coated onto the sample, and the pad areas are opened by UV photolithography.
After lithium tantalate poling, the photoresist, aluminum electrodes, and protective silicon dioxide layer are sequentially removed via wet etching in acetone, aluminum etchant (TechniEtch Al80), and buffered hydrofluoric acid (BHF), respectively.

A hydrogen silsesquioxane (HSQ) layer of approximately \SI{900}{\nano\meter} is then spin-coated to serve as an etching mask. Waveguide patterns are defined using EBL.
After development in a \SI{25}{\percent} tetramethylammonium hydroxide (TMAH) solution, the pattern is transferred into the lithium tantalate layer by argon ion beam etching (Veeco Nexus IBE350).
A precise etch depth is achieved through multi-step etching and calibration, resulting in a slab thickness of 100$\pm$1~nm.
To reduce sidewall roughness caused by redeposition during ion beam etching, a subsequent wet etching step is performed using a high-temperature (\SI{80}{\celsius}) potassium hydroxide (KOH) mixture solution~\cite{wang_LithiumTantalatePhotonic_2024}.
Finally, the PPLT waveguides are clad with a \SI{3.1}{\micro\meter} hydrogen-free SiO$_2$ layer deposited by inductively coupled plasma chemical vapor deposition (ICP-CVD, Oxford PlasmaPro 100)~\cite{qiu_HydrogenfreeLowtemperatureSilica_2024}.
The chip is then singulated through sequential SiO$_2$ dry etching, deep silicon etching, and manual cleavage.

\section{Periodically poled 18 mm-long lithium tantalate waveguides}

We implement periodic poling of the lithium tantalate waveguides using a process flow similar to that described in Ref.~\cite{kuznetsov_WattlevelSecondHarmonic_2025}.
To enable accurate prediction of dispersion and quasi-phase-matching conditions, prior to fabrication, we calculate the optical mode profiles and propagation constants over a broad range of optical frequencies and waveguide cross-sections using a commercially available finite-element-method (FEM) solver, COMSOL Multiphysics\textsuperscript{\textregistered}.

\begin{figure*}[htb]
	\centering
	\includegraphics[width=1\textwidth]{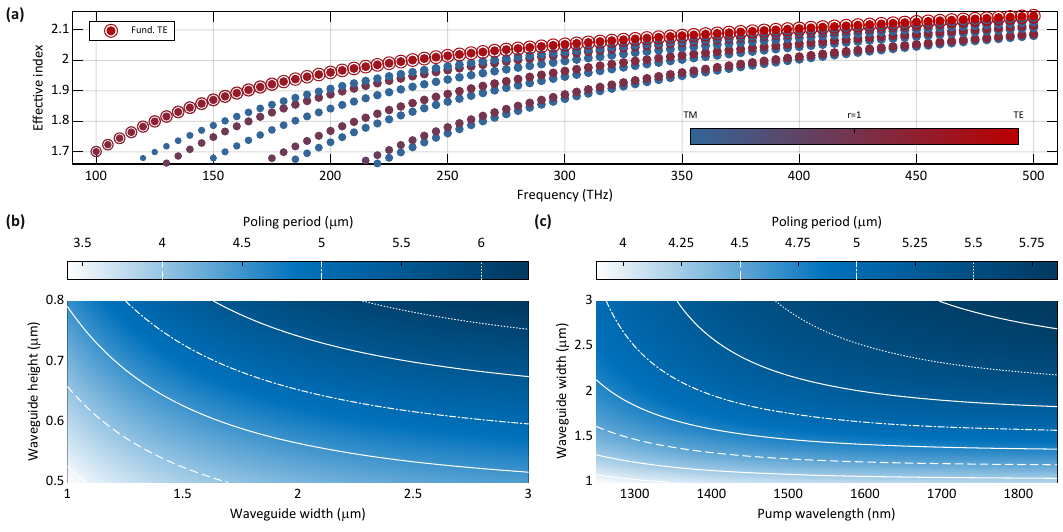}
	\caption{\textbf{Numerical simulations of the propagation constants of optical modes, and calculated poling periods.}
		\textbf{(a)} The first eight optical modes with the highest confinement in the waveguide with the target dimensions.
        The fundamental TE$_{00}$ mode has the strongest confinement, and due to low material birefringence, does not experience mode crossings.
        \textbf{(b)} Poling periods for different waveguide cross-sections; the slab is~\SI{100}{\nano\meter}.
        \textbf{(c)} Poling periods calculated for the pump waves at different frequencies and for the waveguides with different widths; the waveguide height and slab are fixed.
		}
	\label{fig:periodic_poling_design}
\end{figure*}

The relatively low birefringence of lithium tantalate ensures that the fundamental TE$_{00}$ mode remains the most strongly confined mode across the investigated wavelength range, and no mode crossings are observed (Fig.~\ref{fig:periodic_poling_design}(a)).
This simplifies the waveguide design, as the dispersion and phase-matching properties are not perturbed by higher-order mode interactions, and the choice of waveguide dimensions is not constrained by modal degeneracies.

The final waveguide cross-section is selected based on the dispersion engineering considerations described further.
For the chosen geometry, the quasi-phase-matching (QPM) period is calculated assuming a pump wavelength of $\lambda_{p}=\SI{1550}{\nano\meter}$, as shown in Fig.~\ref{fig:periodic_poling_design}(b,c).
To account for the accuracy of our numerical simulations and dimension tolerances, we fabricate waveguides with varied waveguide widths and corresponding poling periods, ensuring that a subset of devices would satisfy the target phase-matching condition after fabrication.
The electrodes for periodic poling are fabricated following the same layout and processing steps as in Ref.~\cite{kuznetsov_WattlevelSecondHarmonic_2025}.
In the present design, the electrode gap is increased to~\SI{20}{\micro\meter} to allow three parallel waveguides to be poled within the same electrode region. 

\begin{figure*}[htb]
	\centering
	\includegraphics[width=1\textwidth]{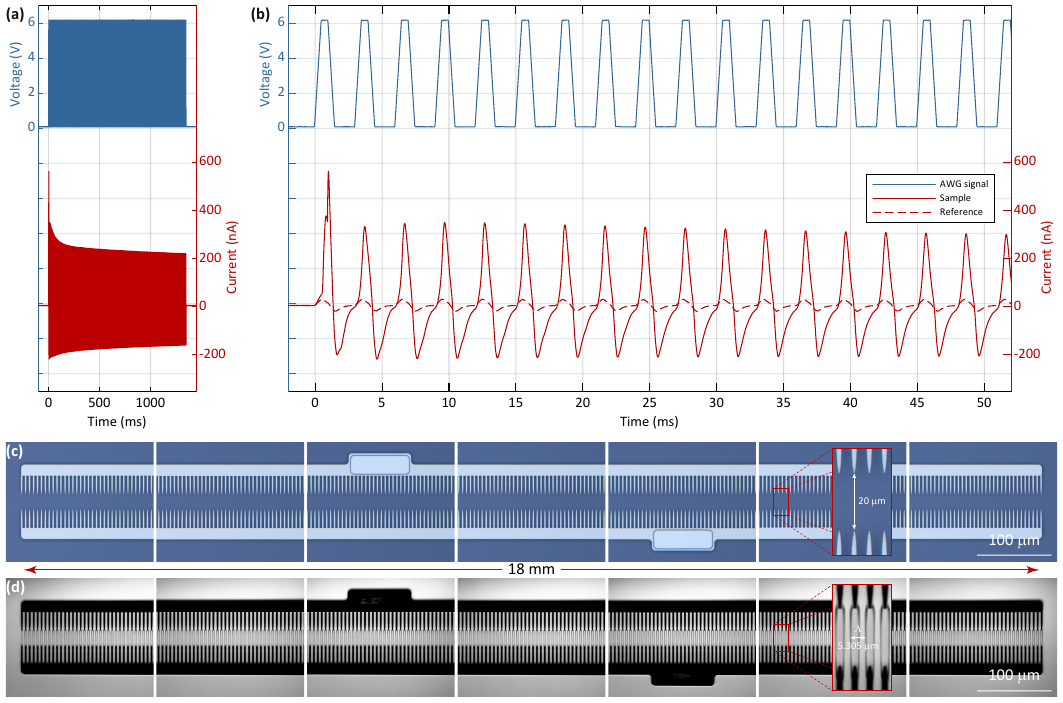}
	\caption{\textbf{Periodic poling of~\SI{18}{\milli\meter}-long lithium tantalate waveguides.}
		\textbf{(a)} The entire poling sequence of~\SI{450}{} pulses, produced by an arbitrary waveform generator (blue), and poling current recorded with an oscilloscope.
        \textbf{(b)} Same as~\textbf{(a)}, but magnified to the beginning of the pulse sequence.
        \textbf{(c)}~Optical microscope images of different segments of~\SI{18}{\milli\meter}-long electrodes before poling, and~\textbf{(d)}~two-photon microscope images of the same electrodes after poling, respectively.
		Scale bars:~\SI{100}{\micro\meter}.
		}
	\label{fig:periodic_poling_pulses}
\end{figure*}

Domain inversion is performed using high-voltage pulses with an amplitude of~\SI{1200}{V} as shown in Fig.~\ref{fig:periodic_poling_pulses}(a,b).
A pulse sequence is produced by an arbitrary waveform generator and then amplified using a high-voltage amplifier with a gain factor of~\SI{200}{}.
A total of~\SI{450}{} pulses are applied, each with~\SI{0.5}{\milli\second} linear rise,~\SI{0.5}{\milli\second} flat top, and~\SI{0.5}{\milli\second} fall segments, separated by~\SI{1.5}{\milli\second} intervals.
The same poling parameters are verified on short (\SI{1}{\milli\meter}) test electrodes prior to application on the full-length devices.
We observe that identical poling conditions yield consistent domain inversion with an approximately~\SI{50}{\percent} duty cycle for both the test structures and the~\SI{18}{\milli\meter}-long electrodes, indicating good process scalability.

The current produced by the charge redistribution process during domain flipping is collected using an oscilloscope~\cite{kuznetsov_WattlevelSecondHarmonic_2025}, and the current trace collected from the poling electrode and the reference trace collected without any samples undergoing poling are shown as solid and dashed red lines, respectively, in Fig.~\ref{fig:periodic_poling_pulses}(b).
Both traces show oscillatory behavior due to stray capacitance in the circuit, and the poling current trace has an asymmetric shape with a decaying envelope, indicating that most of the domain inversion occurs within the first cycles, and subsequent pulses gradually widen the domains.

In the present work, the electrode length is limited by the chip size and the available fabrication equipment; however, we believe that the demonstrated process is compatible with longer electrodes, and we expect that waveguides exceeding~\SI{18}{\milli\meter} in length could be poled using the same fabrication approach on appropriately sized substrates.
The quality and periodicity of the ferroelectric domain inversion were characterized using a scanning two-photon microscope (Zeiss LSM 710 NLO)).
The SHG signal, which is sensitive to the sign of the nonlinear coefficient, provides a direct, non-destructive probe of the domain orientation.
By scanning the focused pump beam along the waveguide, we verified the uniformity of the poling period and confirmed the absence of domain merging or significant duty-cycle distortion over the full~\SI{18}{\milli\meter} length, as shown in Fig.~\ref{fig:periodic_poling_pulses}(d).

\section{Numerical simulations of second-harmonic generation and cascaded parametric amplification}

The nonlinear interactions in the periodically poled lithium tantalate waveguides are modeled using a coupled-mode formalism describing SHG and cascaded OPA processes~\cite{lee_PhaseSensitiveAmplification_2009}.
The simulations assume single-mode propagation for all interacting fields and continuous-wave operation, which is justified by the narrow linewidth of the pump and the absence of rapid temporal dynamics.

First, we calculate propagation constants $\beta_k(\omega_k)$ and transverse mode profiles $f_k(x,y)$ for the pump, second-harmonic, signal, and idler waves using a FEM eigenmode solver.
These quantities determine both the phase-matching conditions and the nonlinear coupling strength through spatial overlap integrals.
The electric field is expanded into guided modes as

\begin{equation}
E(x,y,z,t) = \frac{1}{2} \sum_k f_k(x,y) E_k(z) e^{i(\beta_k z - \omega_k t)} + \mathrm{c.c.}
\end{equation}

Here we introduce $A_k(z)$ -- slowly varying amplitudes normalized such that $|A_k|^2$ represents optical power

\begin{equation}
E_{k}(z) = A_{k}(z) \sqrt{2Z_{k}}, \quad
Z_{k} = \dfrac{1}{\varepsilon_0 c n_{k}} = \dfrac{Z_0}{n_{k}}, \quad
Z_0 = \dfrac{1}{\varepsilon_0 c} = \sqrt{\dfrac{\mu_0}{\varepsilon_0}}, \quad
c = \dfrac{1}{\sqrt{\varepsilon_0 \mu_0}}.
\end{equation}

Here, $ Z_0 $ is the impedance of free space, $ n_{k} $ are effective mode indices, $ c $ is the speed of light, and $ \varepsilon_0 $ and $ \mu_0 $ are the magnetic and electric constants, respectively.
Under the slowly varying envelope approximation and including linear propagation loss $\alpha$, the coupled-mode equations describing SHG are~\cite{lee_PhaseSensitiveAmplification_2009}

\begin{align}
\frac{dA_{\mathrm{p}}}{dz} &= -\frac{\alpha}{2}A_{\mathrm{p}} + i\omega_{\mathrm{p}}\kappa_{\mathrm{shg}} A_{\mathrm{sh}}A_{\mathrm{p}}^{*}e^{-i\Delta\beta_{\mathrm{shg}}^{\mathrm{(eff)}}z}, \\
\frac{dA_{\mathrm{sh}}}{dz} &= -\frac{\alpha}{2}A_{\mathrm{sh}} + i\omega_{\mathrm{p}}\kappa_{\mathrm{shg}}A_{\mathrm{p}}^2 e^{i\Delta\beta_{\mathrm{shg}}^{\mathrm{(eff)}}z}.
\end{align}

For parametric amplification, four interacting waves are considered: pump ($\mathrm{p}$), second harmonic ($\mathrm{sh}$), signal ($\mathrm{s}$), and idler ($\mathrm{i}$).
The coupled-mode equations describing the cascaded SHG and DFG processes are~\cite{lee_PhaseSensitiveAmplification_2009}

\begin{align}
\frac{dA_{\mathrm{p}}}{dz} &= -\frac{\alpha}{2}A_{\mathrm{p}} + i\omega_{\mathrm{p}}\kappa_{\mathrm{shg}} A_{\mathrm{sh}}A_{\mathrm{p}}^{*} e^{-i\Delta\beta_{\mathrm{shg}}^{\mathrm{(eff)}}z}, \\
\frac{dA_{\mathrm{sh}}}{dz} &= -\frac{\alpha}{2}A_{\mathrm{sh}}
+ i\omega_{\mathrm{p}}\kappa_{\mathrm{shg}}A_{\mathrm{p}}^2 e^{i\Delta\beta_{\mathrm{shg}}^{\mathrm{(eff)}}z}
+ 2i\omega_{\mathrm{p}}\kappa_{dfg}A_{\mathrm{s}} A_{\mathrm{i}} e^{i\Delta\beta_{\mathrm{dfg}}^{\mathrm{(eff)}}z}, \\
\frac{dA_{\mathrm{s}}}{dz} &= -\frac{\alpha}{2}A_{\mathrm{s}} + i\omega_{\mathrm{s}}\kappa_{\mathrm{dfg}} A_{\mathrm{sh}}A_{\mathrm{i}}^{*} e^{-i\Delta\beta_{\mathrm{dfg}}^{\mathrm{(eff)}}z}, \\
\frac{dA_{\mathrm{i}}}{dz} &= -\frac{\alpha}{2}A_{\mathrm{i}} + i\omega_{\mathrm{i}}\kappa_{\mathrm{dfg}} A_{\mathrm{sh}}A_{\mathrm{s}}^{*} e^{-i\Delta\beta_{\mathrm{dfg}}^{\mathrm{(eff)}}z}.
\label{eq:opa_coupled_eq}
\end{align}

For SHG and DFG processes, the phase mismatches are defined as

\begin{equation}
\Delta\beta_{\mathrm{shg}} = 2\beta_{\mathrm{p}} - \beta_{\mathrm{sh}}, \quad
\Delta\beta_{\mathrm{dfg}} = \beta_{\mathrm{s}} + \beta_{\mathrm{i}} - \beta_{\mathrm{sh}}.
\end{equation}

In periodically poled waveguides, the QPM condition introduces an additional wavevector $2\pi/\Lambda$, such that the effective mismatches become

\begin{equation}
\Delta\beta_{\mathrm{shg}}^{\mathrm{(eff)}} = 2\beta_{\mathrm{p}} - \beta_{\mathrm{sh}} + \frac{2\pi}{\Lambda}, \quad
\Delta\beta_{\mathrm{dfg}}^{\mathrm{(eff)}} = \beta_{\mathrm{s}} + \beta_{\mathrm{i}} - \beta_{\mathrm{sh}} + \frac{2\pi}{\Lambda}.
\end{equation}

The poling period $\Lambda$ is chosen to satisfy $\Delta\beta_{\mathrm{shg}}^{\mathrm{(eff)}} = 0 $ at the pump wavelength, which simultaneously enables efficient cascaded parametric amplification.
The nonlinear coupling coefficients are determined by the effective second-order nonlinearity and modal overlap.
For periodically poled lithium tantalate, the effective nonlinear coefficient accounts for the first Fourier component of the alternating nonlinearity in the domain-inverted waveguides.
The effective nonlinear coefficient and coupling coefficients for the SHG and DFG processes are given by

\begin{equation}
d_{\mathrm{eff}} = \frac{2}{\pi} d_{33}, \quad
\kappa_{\mathrm{shg}}  = \frac{d_{\mathrm{eff}}}{c}
\sqrt{\frac{{2Z_0}}{{n_{\mathrm{p}}^2 n_{\mathrm{sh}}}}}
f_{0_{\mathrm{shg}}},\quad
\kappa_{\mathrm{dfg}}  = \frac{d_{\mathrm{eff}}}{c}
\sqrt{\frac{{2Z_0}}{{n_{\mathrm{s}} n_{\mathrm{i}} n_{\mathrm{sh}}}}}
f_{0_{\mathrm{dfg}}}.
\end{equation}

The spatial overlap integral between interacting modes is defined as

\begin{equation}
f_{0_{\mathrm{shg}}} = \iint f_{\mathrm{sh}}(x,y) f_{\mathrm{s}}(x,y) f_{\mathrm{i}}(x,y)\,dx\,dy, \quad
f_{0_{\mathrm{dfg}}} = \iint f_{\mathrm{sh}}(x,y) (f_{\mathrm{p}}(x,y))^2 \,dx\,dy.
\end{equation}

The mode profiles are normalized such that $\iint |f_k|^2 dxdy = 1$.
The magnitudes of overlap integrals depend strongly on the waveguide cross-section and is calculated directly from the mode fields obtained in the FEM solver.
The coupled-mode equations are solved numerically along the propagation coordinate using a forward-Euler integration scheme.
The input conditions are specified as

\begin{equation}
A_{\mathrm{p}}(0) = \sqrt{P_{\mathrm{p}_{\mathrm{in}}}}, \quad
A_{\mathrm{sh}}(0) = 0, \quad
A_{\mathrm{s}}(0) = \sqrt{P_{\mathrm{s}_{\mathrm{in}}}}, \quad
A_{\mathrm{i}}(0) = 0.
\end{equation}

\begin{figure*}[htb]
	\centering
	\includegraphics[width=1\textwidth]{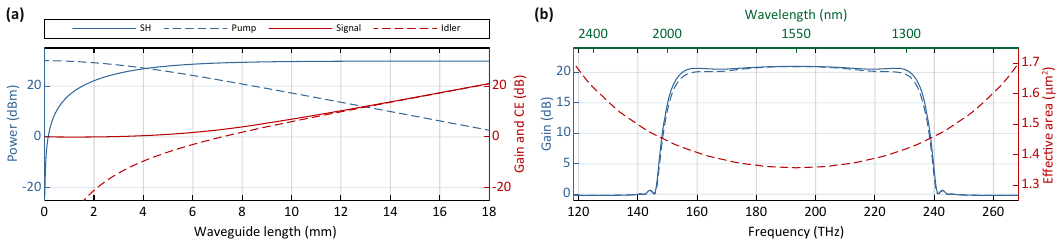}
	\caption{\textbf{Numerical simulation of second-harmonic generation and cascaded parametric amplification.}
		\textbf{(a)} Four quasi-phase-matched waves propagating along the lithium tantalate waveguide with second-order nonlinearity.
        \textbf{(s)} Mode overlap contribution for far-separated interacting waves.
		}
	\label{fig:simulations}
\end{figure*}

The examples of the results of numerical simulations are shown in Fig.~\ref{fig:simulations}.
We set the pump power to~\SI{1}{\watt}, and the signal power to~\SI{1}{\micro\watt} to keep the pump undepleted, and use these numbers for all simulation results presented here and further.
As the calculation of the mode overlap is time-consuming, we simplify the model by noting that the mode overlap between the remote signal and the idler waves is, in fact, nearly the same as the overlap between the pump and SH waves.
The equations can be simplified by approximating the model overlap integrals as

\begin{equation}
f_{0_{\mathrm{shg}}} = f_{0_{\mathrm{dfg}}} = f_{0} = \sqrt{\dfrac{1}{A_{\mathrm{eff}}}},
\end{equation}

where $ A_{\mathrm{eff}} $ is the effective mode area.
Fig.~\ref{fig:simulations}(b) confirms that this approximation is valid, and we use it in all our simulations.
The normalized SHG efficiency is defined using the input pump and the output SH powers and can also be expressed using the SHG coupling coefficient.

\begin{equation}
\eta = \dfrac{P_{\mathrm{sh}}(L)}{P_{\mathrm{p}}(0)^2L^2}\cdot100\%, \quad
\eta = \kappa_{\mathrm{shg}}^2\omega_{\mathrm{p}}^2.
\end{equation}

To keep our simulation results conservative and avoid overestimating the gain, we set the SHG efficiency to~\SI{515}{\percent\watt^{-1}\centi\meter^{-2}}, which corresponds to the material nonlinearity d$_{33}$
equal to~\SI{10}{\pico\meter\volt^{-1}} -- in fact, slightly less than the value reported in Ref.~\cite{shoji_AbsoluteScaleSecondorder_1997}, but closer to our experimental results.

\section{Dispersion engineering for broadband amplification}

The bandwidth of parametric amplification is determined by the phase mismatch between the interacting waves.
In cascaded second-order nonlinear processes, periodic poling ensures QPM condition of the individual second-order interactions but does not directly set the amplification bandwidth, which is instead governed by the waveguide dispersion.

In a periodically poled waveguide, the SHG and DFG processes satisfy the QPM conditions

\begin{equation}
2\beta_{\mathrm{p}} - \beta_{\mathrm{sh}} + \frac{2\pi}{\Lambda} = 0, \quad
\beta_{\mathrm{s}} + \beta_{\mathrm{i}} - \beta_{\mathrm{sh}} + \frac{2\pi}{\Lambda} = 0.
\label{eq:qpm_shg}
\end{equation}

Combining both equations to eliminate the second-harmonic term yields the effective phase-matching condition for cascaded parametric amplification,

\begin{equation}
\Delta \beta_{\mathrm{opa}} =  2\beta_{\mathrm{p}} - \beta_{\mathrm{s}} - \beta_{\mathrm{i}} = 0,
\label{eq:effective_pm}
\end{equation}
which is formally identical to the phase-matching condition of degenerate four-wave mixing.
This equivalence allows the cascaded $\chi^{(2)}$ interaction to be treated using the same dispersion formalism as Kerr-based parametric processes.

\begin{figure*}[htb]
	\centering
	\includegraphics[width=1\textwidth]{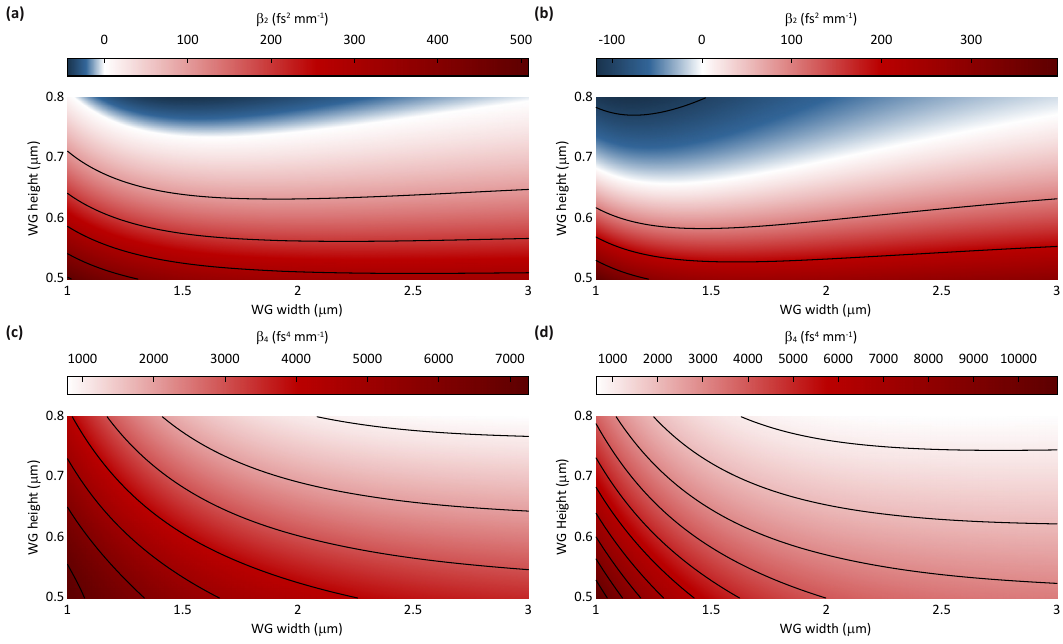}
	\caption{\textbf{Numerically calculated dispersion maps for the waveguides with different dimensions.}
		\textbf{(a,c)} Second- and fourth-order dispersion coefficients for the waveguides with a~\SI{300}{\nano\meter}-thick slab.
        \textbf{(b,s)} Second- and fourth-order dispersion coefficients for the waveguides with a~\SI{100}{\nano\meter}-thick slab.
        A smaller slab allows anomalous dispersion to be achieved at a smaller total waveguide height.
		}
	\label{fig:dispersion_maps}
\end{figure*}

Expanding Eq.~(\ref{eq:effective_pm}) in a Taylor series around the pump frequency $\omega_{\mathrm{p}}$ gives

\begin{equation}
\Delta \beta_{\mathrm{opa}} \approx \beta_{2}\Omega^{2} + \frac{1}{12}\beta_{4}\Omega^{4} + \cdots,
\label{eq:taylor_pm}
\end{equation}
where $\Omega = \omega_{\mathrm{p}} - \omega_{\mathrm{s}}$ is the signal frequency detuning from the pump, and $\beta_{2}$ and $\beta_{4}$ are the second- and fourth-order dispersion coefficients evaluated at $\omega_{\mathrm{p}}$.
In the dual-pump configuration employing SFG instead of SHG, the same formalism applies, with $\beta_{2}$ and $\beta_{4}$ corresponding to an effective virtual pump located midway between the two physical pump frequencies.
The equation describing the SHG part in Eqs.~(\ref{eq:opa_coupled_eq}) should be replaced by another set of equations describing the DFG part, with corrected input conditions. 
The gain bandwidth is therefore primarily determined by the dispersion coefficients of the waveguide.
By tailoring the waveguide cross-section, $\beta_{2}$ and $\beta_{4}$ can be engineered to achieve broadband amplification (Fig.~\ref{fig:dispersion_maps}).
In contrast to Kerr-based parametric amplifiers, the contribution of nonlinear phase shifts due to self- and cross-phase modulation is negligible in the present devices because of the short interaction length and the relatively low Kerr nonlinearity of lithium tantalate; this is confirmed by the experimental data presented in the main text, as the gain profiles do not feature Kerr-like gain lobes.
Consequently, the gain bandwidth is governed predominantly by the second- and fourth-order dispersion terms.
While reduced normal $\beta_{2}$ can provide moderate bandwidth broadening, a more effective strategy is to operate in a regime with low anomalous $\beta_{2}$ and normal $\beta_{4}$.
In this case, the fourth-order dispersion compensates the phase mismatch at large detuning, leading to the formation of two remote gain bands.
With appropriate dispersion engineering, these sidebands can merge with the central gain band, producing a broad and top-flat gain spectrum.
To identify suitable waveguide geometries, we compute dispersion maps of $\beta_{2}$ and $\beta_{4}$ as functions of waveguide width and total film thickness for slab thicknesses of~\SI{300}{\nano\meter} and~\SI{100}{\nano\meter}, shown in Fig.~\ref{fig:dispersion_maps}, in left panels and right panels, respectively.
The thicker slab requires a thicker starting substrate to reach the anomalous dispersion regime.
For a~\SI{100}{\nano\meter} slab, anomalous $\beta_{2}$ is obtained for total film thicknesses exceeding approximately~\SI{680}{\nano\meter}.

\begin{figure*}[htb]
	\centering
	\includegraphics[width=1\textwidth]{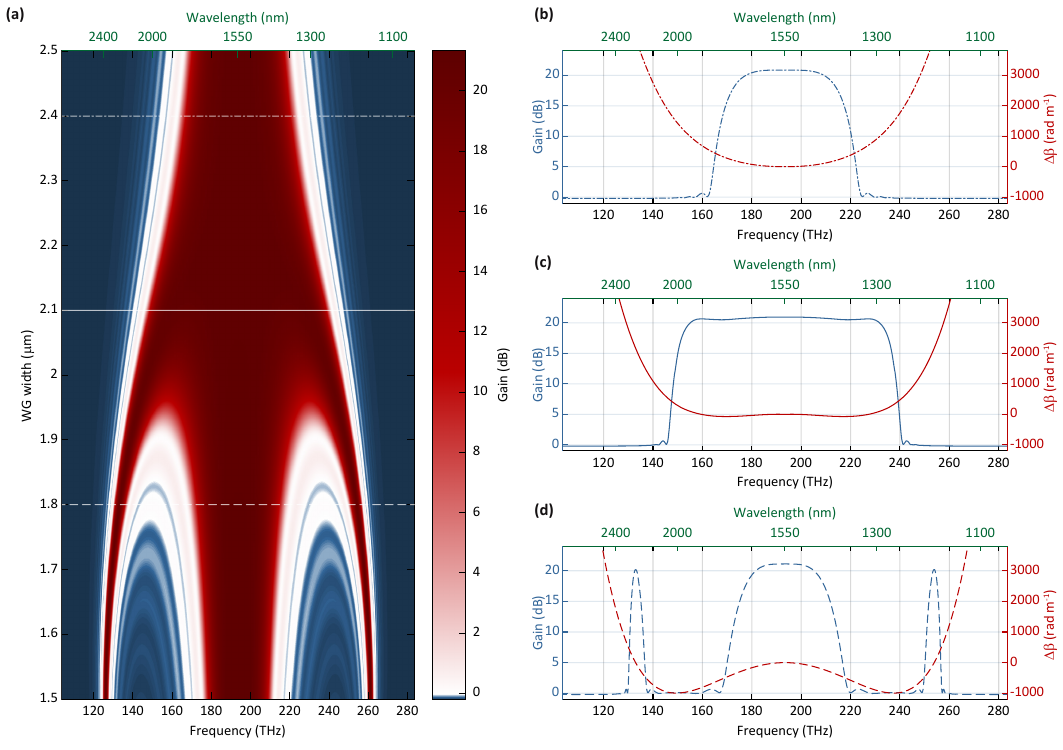}
	\caption{\textbf{Parametric gain for different waveguide widths.}
		\textbf{(a)} Full gain map, calculated for the fixed slab thickness and total waveguide height, and varied waveguide width.
        Panels \textbf{(a--c)} show slices of the map for normal, near-zero anomalous, and strongly anomalous dispersion regimes, respectively.
        }
	\label{fig:gain_vs_width}
\end{figure*}

As we explain in the main text, there is a certain deviation between our numerical simulation and experimental results, attributed to the inaccuracy of the refractive index data of bulk crystals applied to thin-film material and to the deviations of the idealized models from the real waveguide geometry.
We compensate for this offset by setting the effective waveguide height to~\SI{712}{\nano\meter} and using this value for all numerical simulations described in this work.
The calculated dispersion shows a strong dependence on waveguide height and moderate dependence on waveguide width, enabling fine lithographic tuning of the dispersion at fixed film thickness.
This allows multiple waveguides with different widths to be fabricated on the same chip, with post-fabrication selection of the geometry providing the desired dispersion profile.
Figure~\ref{fig:gain_vs_width}(a) presents simulated gain spectra as a function of waveguide width.

For wider waveguides, where $\beta_{2}$ is strongly normal, the gain bandwidth is limited but remains single-peaked.
As the width decreases and $\beta_{2}$ becomes increasingly anomalous, two gain sidebands emerge and shift further away from the central band.
At a waveguide width of approximately~\SI{2.1}{\micro\meter}, the second-order dispersion is anomalous but sufficiently small for the sidebands to merge with the central band.
This condition results in an exceptionally wide and flat gain spectrum, which is used as the operating point in the main text.

\section{Fabrication tolerance of the waveguide cross-section dimensions}

\begin{figure*}[htb]
	\centering
	\includegraphics[width=1\textwidth]{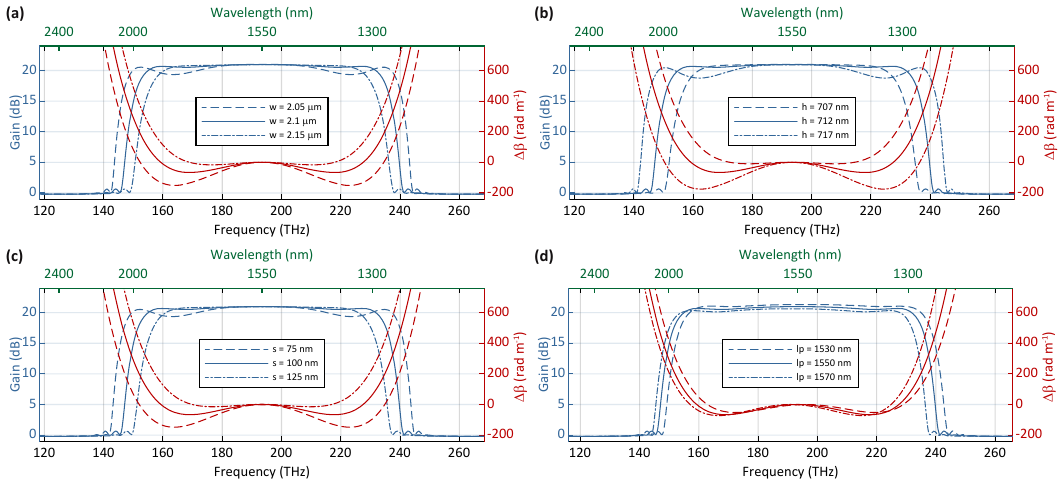}
	\caption{\textbf{Tolerance of the parametric gain profile to the variations of the waveguide dimensions and variations of the pump wavelength.}
		\textbf{(a--c)} Waveguide width, height, and slab are varied by~\SI{50}{\nano\meter},~\SI{5}{\nano\meter}, and~\SI{20}{\nano\meter}, respectively, around optimal design dimensions.
        \textbf{(d)} Pump wavelength is varied by~\SI{20}{\nano\meter}, assuming that the QPM condition is satisfied perfectly in each case.
		}
	\label{fig:fabrication_tolerance}
\end{figure*}

Fabrication-induced deviations in waveguide geometry, arising from lithography and etching processes, can modify the phase-matching condition and therefore affect the gain bandwidth and spectral profile of the OPA.
We numerically simulate the parametric gain while independently varying the waveguide width, total height, and slab thickness, keeping all other parameters fixed at their nominal design values.
The resulting gain spectra are shown in Fig.~\ref{fig:fabrication_tolerance}(a--c) for width, height, and slab variations, respectively.

The simulations indicate that the proposed waveguide design exhibits high robustness to typical fabrication imperfections.
Variations of the waveguide width of up to~\SI{50}{\nano\meter} produce only minor changes in the gain bandwidth and spectral shape, relaxing lithographic resolution requirements and enabling the use of lower-cost fabrication processes for large-scale production.
However, high tolerance comes at a cost of significant changes in the waveguide design in case of inaccuracies in the model -- that is why the final target width of PPLT waveguides in this work is~\SI{2.1}{\micro\meter} instead of~\SI{1.8}{\micro\meter}.
In fact, the refractive index data we use~\cite{moutzouris_TemperaturedependentVisibleNearinfrared_2011} belongs to the Mg:LiTaO$_3$ compound.
Although the material birefringence is low, both ordinary and extraordinary material indices have to be considered in numerical simulations.
Among recently published works, Ref.~\cite{moutzouris_TemperaturedependentVisibleNearinfrared_2011} appears to be the most relevant as it shows both indices measured using the same methods and instruments across the wavelength range relevant to our work.
Other sources we are aware of either do not provide data for both indices or focus on a different wavelength window.
Recent works might improve the precision of numerical models.
In contrast, the total waveguide height has the strongest impact on the gain spectrum.
Acceptable performance is maintained for height deviations within approximately~\SI{5}{\nano\meter}.
Such tolerances are compatible with standard wafer fabrication processes and remain less stringent than those typically required for long Kerr-based nonlinear waveguides, where phase matching is more sensitive to thickness variations.
The slab thickness shows intermediate sensitivity, with variations of up to~\SI{20}{\nano\meter} having a limited effect on the gain characteristics.
This relaxes constraints on partial etching steps and reduces the need for tight end-point control during slab definition.
In addition to geometrical tolerances, we evaluate the sensitivity of the device to deviations in the pump wavelength.
Since the poling period is defined before waveguide fabrication, fabrication-induced dispersion shifts can lead to an offset between the designed and actual phase-matching wavelengths.
Figure~\ref{fig:fabrication_tolerance}(d) shows that the gain spectrum remains largely unchanged for pump wavelength detuning of up to~\SI{20}{\nano\meter}.
This tolerance facilitates practical operation within the optical C-band, where high-power pump sources are readily available.
Overall, the high tolerance of the presented OPA design to both geometrical and pump wavelength variations simplifies the fabrication workflow and is expected to improve device yield and reproducibility.

\section{Parametric fluorescence -- extended data}

\begin{figure*}[htb]
	\centering
	\includegraphics[width=1\textwidth]{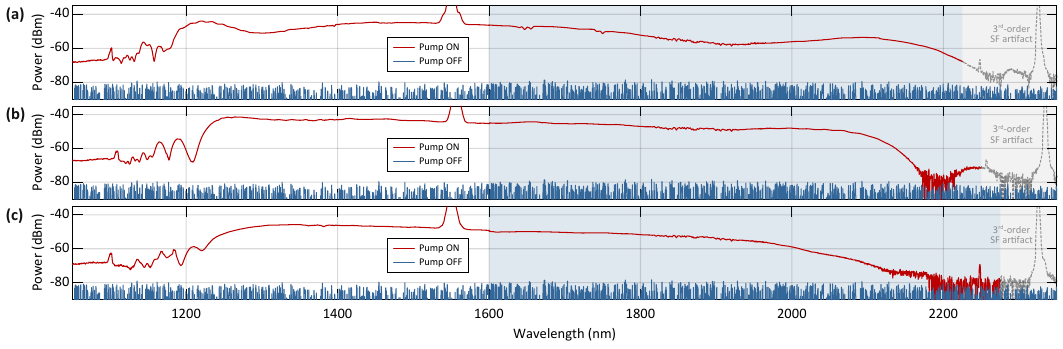}
	\caption{\textbf{Parametric fluorescence spectra in waveguides with different widths.}
		\textbf{(a--c)} Waveguides with designed widths of~\SI{2.05}{\micro\meter}, \SI{2.1}{\micro\meter}, and \SI{2.15}{\micro\meter}, respectively.
        The waveguide with a spectrum shown in \textbf{(b)} is the waveguide discussed in the main text.
       }
	\label{fig:fluo_extended}
\end{figure*}

Fig.~\ref{fig:fluo_extended} shows additional measurements of parametric fluorescence spectra.
The transition from the anomalous (Fig.~\ref{fig:fluo_extended}(a)) to the normal dispersion regime (Fig.~\ref{fig:fluo_extended}(b)) is clearly marked by the merging of the gain sidebands with the central gain region.

\section{Interband modulation transfer}

\begin{figure*}[htb]
	\centering
	\includegraphics[width=1\textwidth]{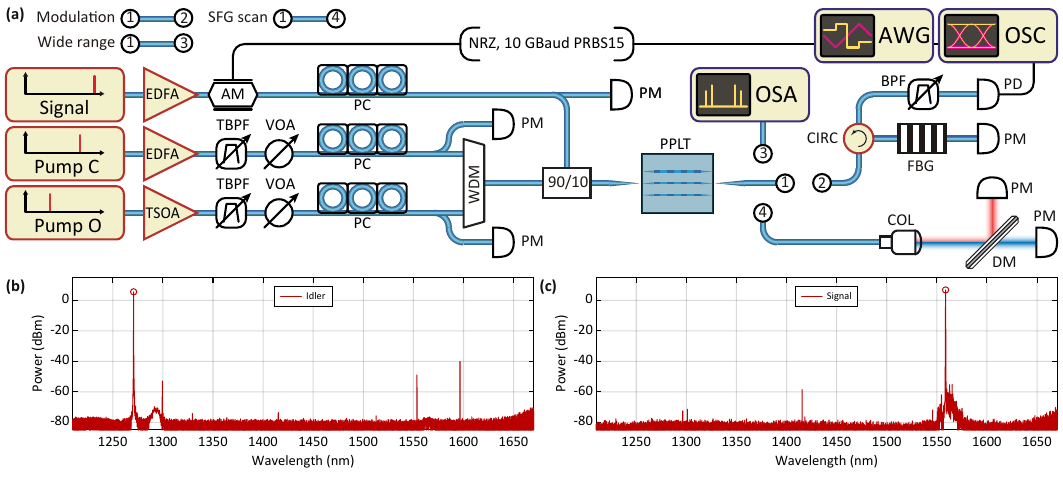}
	\caption{\textbf{Optical modulation transfer experiment.}
		\textbf{(a)}~A schematic of the experimental setup for sum-frequency measurements and telecom-datacom modulation transfer.
		TSOA: tapered semiconductor optical amplifier.
		EDFA: erbium-doped fiber amplifier.
		TBPF: tunable bandpass filter.
		VOA: variable optical attenuator.
		AM: amplitude modulator.
		FBG: fiber Bragg grating.
		PD: photodiode.
		WDM: wavelength-division multiplexer.
		AWG: arbitrary waveform generator.
		PC: polarization controller.
		PM: power meter.
		COL: collimator.
		DM: dichroic mirror.
		CIRC: circulator.
		OSA: optical spectrum analyzer.
		OSC: oscilloscope.
		\textbf{(b)} Idler at~\SI{1271}{\nano\meter}, generated by the signal at~\SI{1596}{\nano\meter}; the received optical power is~\SI{5.7}{\dBm}.
        \textbf{(c)} Signal at~\SI{1559}{\nano\meter}; the received optical power is~\SI{6.9}{\dBm}.
        }
	\label{fig:modulation}
\end{figure*}

The main challenge in the modulation transfer experiment is to suppress residual pump power, ensure that only the relevant tones are received by the photodiode, and maintain sufficient optical power for detection (Fig.~\ref{fig:modulation}(a)).

As explained in the main text, we use a signal at~\SI{1559}{\nano\meter} for the reference measurements before the chip and for the amplified signal measurements after the chip.
For the idler measurements, we set the signal wavelength to~\SI{1596}{\nano\meter}, generating the idler wave at~\SI{1271}{\nano\meter}.
For both signal and idler measurements, we use available FBGs and circulators to isolate the desired tones.
However, this approach does not provide sufficient suppression of unwanted signals, and we therefore employ an additional WDM and FBG to suppress residual pumps.

Despite limited pump power in the O-band and~\SI{2.5}{\dB} of losses in the filtering stage, we are able to deliver sufficient output power to the photodiode, as shown in Fig.~\ref{fig:modulation}.
In this experiment, the on-off gain is relatively low at~\SI{3.5}{\dB}, because the PPLT OPA operates in the saturation regime -- we use~\SI{10}{\dBm} of input signal power to ensure that sufficient idler power is delivered to the photodiode.
We further amplify the electrical signal from the photodiode using a broadband RF amplifier (iXblue DR-AN-28-MO).

Although the photodiode is rated for dual-band operation, its responsivity in the O-band is lower.
According to the datasheet, the responsivity at~\SI{1271}{\nano\meter} is nearly twice as low as in the C-band.
This contributes to the degradation of the eye diagram for the idler measurement shown in the main text.

\bibliography{bibliography}